\documentclass[a4paper,11pt]{article}
\usepackage{jheppub} % for details on the use of the package, please see the JINST-author-manual
\usepackage{lineno}
\usepackage{braket}
\usepackage{slashed}
\usepackage{bbold}
\usepackage{amsmath}
\usepackage{mathtools}
\allowdisplaybreaks

\newcommand{\dd}{\mathrm{d}}
\newcommand{\be}{\begin{equation}}
\newcommand{\ee}{\end{equation}}

\newcommand{\nn}{\text}

\title{\boldmath D-terms in Generalised Complex Geometry}

\author{Vincent Menet}
\affiliation{Laboratoire de Physique Théorique et Hautes Energies - LPTHE\\
Sorbonne Université, 4 Place Jussieu, 75005 Paris, France}

\emailAdd{vmenet@lpthe.jussieu.fr}
\abstract{Generalised Complex Geometry provides a natural interpretation of the $\mathcal{N}=1$ supersymmetry conditions for warped solutions of type II supergravity as differential equations on polyforms on the internal manifold. Written in this language the supersymmetry conditions correspond to  calibration conditions for probe D-branes:  D-string, domain-wall or space-filling
branes, depending on the directions they span in the non-compact  four-dimensional space. The BPS condition corresponding to the calibration of space-filling D-branes has been reformulated by Tomasiello, eliminating the explicit dependence on the metric. We generalise this derivation to the case of non-supersymmetric backgrounds violating the domain-wall and D-string calibration conditions. We use this reformulation to derive constraints that the ten-dimensional solutions with BPS space-filling sources must respect in order to dimensionally reduce to solutions of four-dimensional $\mathcal{N}=1$ supergravity with non-vanishing F-terms and potentially non-vanishing D-terms. We give the equations of motion for the class of type II vacua satisfying these constraints in the language of pure spinors. We investigate how restrictive these constraints are for the class of type IIB SU$(3)$ backgrounds with BPS space-filling O5-planes.}

\begin{document}
\maketitle
\flushbottom
\newpage \section{Introduction}

    Constructing supergravity solutions with D-terms is difficult. Indeed, the prototypical example of D-term supersymmetry breaking is the Fayet–Iliopoulos term, and realising its (field-dependent) embedding in supergravity at energy parametrically lower than the Planck scale turns out to be challenging \cite{Komargodski:2009pc, Dine:1987xk}.

    In this paper, we investigate the possibility of having supergravity solutions with D-terms from a different angle, using  the framework of Generalised Complex Geometry (GCG).
    We consider solutions that are warped products of four-dimensional Minkowski space and an internal six-dimensional compact manifold.  In GCG, the supersymmetry conditions for $\mathcal{N}=1$ vacua
    can be recast in a set of three differential equations on polyforms defined only on the internal compactification space \cite{Grana:2005sn}, called the pure spinors. Moreover, each of these three conditions have an interpretation in terms of calibration conditions for different probe D-branes in the geometry \cite{Martucci:2005ht}: branes filling all the external space (space-filling), and branes that are domain-wall or string-like in the external space.

In this paper we break supersymmetry in a sort of controlled way: we consider first order differential equations which correspond to deformations of the pure spinor equations via the introduction of supersymmetry breaking terms, and then we impose some further requirements from the equations of motion which guarantee that we have actual solutions of supergravity.

 We focus on a class of vacua that still admit BPS space-filling sources. This means that the BPS condition associated to the calibration condition of space-filling D-branes, dubbed the gauge BPSness condition in \cite{Lust:2008zd}, is preserved, while we allow for the conditions corresponding to the calibrations of string-like and domain-wall probe D-branes to be violated. 
        
        In the supersymmetric case, the gauge BPSness condition has been reformulated in \cite{Tomasiello:2007zq}, eliminating the explicit metric dependence, and introducing a generalised version of the Dolbeault operator. We will generalise this derivation to the case of non-supersymmetric vacua violating the string and domain-wall BPSness conditions.

         Still in the $\mathcal{N}=1$ case, the gauge and domain-wall BPSness conditions have been identified with F-term conditions, while the string BPSness has been interpreted as a D-term condition for the dimensionally reduced four-dimensional $\mathcal{N}=1$ theory in \cite{Koerber:2007xk, Martucci:2006ij}. 

         Within our class of non-supersymmetric backgrounds preserving the gauge BPSness, we will focus on those who can be dimensionally reduced to four-dimensional $\mathcal{N}=1$ supergravity solutions with non-vanishing F-terms, and possibly non-vanishing D-terms. To do so, we require that our set of modified pure spinor equations continues to have an interpretation in terms of either (D-) F-term or (D-) F-term conditions. In particular, the gauge BPSness should still be identified with an F-term condition. 
         
         Interestingly, this procedure constrains some supersymmetry breaking terms entering the modified D-string calibration condition, and therefore the possible D-terms. More precisely, we will see that the supersymmetry breaking terms set to zero by this requirement belong to vector representations of the SU$(3)\times$ SU$(3)$ structure defined by the pure spinors\footnote{See the discussion in the text.}, just like the massive spin $\frac{3}{2}$-multiplet degrees of freedom of the four-dimensional theory \cite{Grana:2009im,Grana:2011nb}.

         On another note, we will give the effective potential for our class of backgrounds, written as an integral over the internal compactification space, in the language of pure spinors. We will see that the requirement to interpret the gauge BPSness condition as an F-term condition results in the vanishing of some negative semi-definite contributions to the effective potential, which naturally fits within the four-dimensional $\mathcal{N}=1$ picture, given that these contributions do not originate from the superpotential.

         We derive the equations of motion for our class of backgrounds dimensionally reducing to four-dimensional $\mathcal{N}=1$ supergravity solutions with non-vanishing F-terms and D-terms, and for a subclass where the D-term contributions to the effective potential are set to zero on-shell, restoring the D-string BPSness. Solutions of this kind correspond to vacua with only F-terms, and include the amply discussed class of no-scale vacua. The relative simplicity of these equations of motion compared to the general ten-dimensional type II equations of motion could be used to find new non-supersymmetric flux vacua.

         Finally, the requirement to interpret the gauge BPSness condition as an F-term condition turns out to be rather restrictive on the possible D-terms, and we investigate this in a more concrete way for a class of SU$(3)$ backgrounds with parallel BPS O5-planes.

         The outline of the paper is as follows. In Section \ref{sec:GCG}, we briefly recall how the $\mathcal{N}=1$ supersymmetry conditions for warped compactifications are recast in the language of Generalised Complex Geometry and how ordinary calibrations can be extended to this framework to describe supersymmetric D-brane probes in these backgrounds. We then reformulate the calibration condition associated to space-filling sources in the non-supersymmetric case. In Section \ref{sec:effpot}, we briefly review the four-dimensional $\mathcal{N}=1$ interpretation of the BPS conditions in the GCG framework, and we introduce the general type II effective potential in the language of pure spinors. In Section \ref{sec:n0}, we derive the constraint that the gauge BPSness should obey in order to be interpreted as an F-term condition, and we derive the equations of motion for backgrounds satisfying this constraint. In Section \ref{sec:example}, we apply this constraint to SU$(3)$ backgrounds with parallel BPS O5-planes, and study the restriction it imposes on their possible D-terms.
    
\section{Flux vacua in Generalised Complex Geometry}\label{sec:GCG}

\subsection{Compactification and pure spinors}
   We consider type II supergravity backgrounds that are the warped product of four-dimensional Minkowski space $X_4$ and a six-dimensional compact manifold $\mathcal{M}$, with the following metric ansatz
\be 
\label{10dmet}
\dd s^2_{10}=e^{2A(y)}\eta_{\mu\nu}\dd x^\mu\dd x^\nu+g_{mn}\dd y^m\dd y^n,
\ee
where $x^\mu$, $\mu=0,...,3$ are the external coordinates on $X_4$, and $y^m$, $m=1,...,6$ are the coordinates on $\mathcal{M}$. 

The Poincar\'e invariance of $X_4$  constrains the  NSNS and RR-fluxes: the NSNS-field-strength $H$ can only have internal legs, and the ten-dimensional RR-field-strength must take the form\footnote{We use the  democratic formulation of \cite{Bergshoeff:2001pv}, where 
$$F^{10}=\sum_q F^{10}_q $$
with $q=0,2,...10 $ for type IIA and $q=1,3,...9$ for type IIB. }
\be F^{10}=F+e^{4A}\text{vol}_4\wedge\Tilde{F},\label{RR},\ee
where $F$ and $\Tilde{F}$ are purely internal and are related by the self duality of $F^{10}$ as
    \begin{equation}
         \Tilde{F}=\Tilde{\ast}_6 F = \ast_6 \sigma (F),\label{fselfdual}
    \end{equation} 
with $\sigma$ the reversal of all form indices. 

\medskip

In this Section  we focus on backgrounds preserving $\mathcal{N}=1$ supersymmetry,  which amounts to the vanishing of the following gravitino and dilatino variations
  \begin{align}
  \label{10dsusyg}
\delta_\epsilon\psi_M=\left(\nabla_M+\frac{1}{4}\iota_M\slashed{H}\sigma_3+\frac{e^\phi}{16} \begin{pmatrix}
    0 & \slashed{F}^{10}\\
    -\sigma(\slashed{F}^{10}) & 0
\end{pmatrix}\Gamma_M\Gamma_{(10)}\right) \begin{pmatrix}
           \epsilon_1 \\
          \epsilon_{2}
         \end{pmatrix}\\
  \label{10dsusyd}
\delta_\epsilon\lambda=\left(\slashed{\partial}\phi+\frac{1}{2}\slashed{H}\sigma_3+\frac{e^\phi}{16} \Gamma^M\begin{pmatrix}
    0 & \slashed{F}^{10}\\
    -\sigma(\slashed{F}^{10}) & 0
\end{pmatrix}\Gamma_M\Gamma_{(10)}\right)\begin{pmatrix}
           \epsilon_1 \\
          \epsilon_{2}
         \end{pmatrix},
  \end{align}
where $\epsilon_1$ and $\epsilon_2$ are ten-dimensional Majorana-Weyl  spinors, and for a $p$-form $\omega$ the slash symbol denotes
\begin{equation}
\slashed{\omega}=\frac{1}{p!}\omega_{M_1...M_p}\Gamma^{M_1...M_p}. 
\end{equation}

\medskip
    
For backgrounds of warped type the spinors $\epsilon_1$ and $\epsilon_2$ split in the following way
  \begin{equation}
\epsilon_1=\zeta\otimes\eta_{1}+c.c.\qquad \epsilon_2=\zeta\otimes\eta_{2} +c.c.\label{10dspinors}
\end{equation}
where $\zeta$ is a Weyl spinor of positive chirality on $X_4$, and 
$\eta_{1}$ and $\eta_{2}$ are Weyl spinors on the six-dimensional internal space. $\eta_{1}$ has positive chirality, while  $\eta_2$
has negative chirality in type IIA and positive chirality in type IIB.  
In the rest of the paper we will assume that the spinors $\eta_1$ and $\eta_2$ are globally defined such that each define an SU$(3)$ structure on $\mathcal{M}$.

\medskip

The vanishing of the supersymmetry variations  \eqref{10dsusyg}, \eqref{10dsusyd} is equivalent to the set of differential equations \cite{Grana:2005sn}
\begin{align}
  \dd_H(\text{e}^{3A-\phi}\Psi_2)=& 0 \label{ret}\\ 
  \dd_H(\text{e}^{2A-\phi}\text{Im}\Psi_1)=&\ 0\label{second}\\
   \dd_H(e^{4A-\phi}\text{Re}\Psi_1)=&\ e^{4A}\Tilde{\ast}_6 F\label{susy3}
\end{align}
where $\dd_H=\dd+H\wedge$, and  $\Psi_1$ and $\Psi_2$ are polyforms 
defined from the internal supersymmetry parameters\footnote{Strictly speaking one should think of these tensor product in terms of the following Fierz identity \be \eta\otimes\chi=\sum_{k=0}^6 \frac{1}{k!}\left(\chi^\dagger\gamma_{m_k...m_1}\eta\right)\gamma^{m_1...m_k}.\ee These tensor products are then isomorphic to polyforms through the Clifford map \eqref{cliffmap}, so we treat them as such from now on. The conventions for the internal gamma matrices are given in Appendix \ref{sec:AppA}.}
\begin{align}
\label{puresp1}
        \Psi_1=&-\frac{8i}{\lVert \eta \rVert^2}\eta_1\otimes\eta_2^\dagger\\
  \label{puresp2}      
         \Psi_2=&-\frac{8i}{\lVert \eta \rVert^2}\eta_1\otimes\eta_2^T.
    \end{align}
The polyforms $\Psi_1$ and $\Psi_2$ are odd/even and even/odd in type IIA/IIB, respectively
\be \Psi_1=\Psi_\mp,\quad \Psi_2=\Psi_\pm \, . \ee

The previous equations have a nice interpretation  within the framework of Generalised Complex Geometry (GCG). 
In Generalised Complex Geometry, one replaces the tangent bundle of the internal manifold with a generalised tangent bundle \cite{Hitchin:2003cxu} 
\begin{equation}
    E\simeq T\mathcal{M}\oplus T^\ast\mathcal{M} \, , 
\end{equation} 
whose sections are locally sums of vectors and one-forms (see Appendix \ref{sec:AppB} for more details)
\begin{equation}
   V=v+\xi\in\Gamma(E).
\end{equation}
In this language, the polyforms $\Psi_1$ and $\Psi_2$ are interpreted as ${\rm Cliff(6,6)}$ pure spinors, which, by construction, are globally defined and compatible
\begin{align}
&\braket{\Psi_1,V\cdot\Psi_2}=\braket{\Psi_1,V\cdot\bar{\Psi}_2}=0\quad\forall \ V\in E\\
&\braket{\Psi_1,\bar{\Psi}_1}=\braket{\Psi_2,\bar{\Psi}_2}=-8i\text{vol}_6,
    \end{align}
where $\braket{\ ,\ }$ denotes the Mukai pairing 
\begin{equation}
\braket{\omega,\chi}=\omega\wedge\sigma(\chi)|_6,\label{Mukai}
\end{equation}
and $\cdot$ is the Clifford action of the generalised vector $V$  on generalised spinors defined in \eqref{vecspinoract}.

The natural inner product between two generalised vectors $V=v+\xi\in\Gamma(E)$ and $W=  w+\rho\in\Gamma(E)$  
\be 
   (V, W)\coloneqq\iota_v\rho+\iota_w\xi
   \ee
defines an SO$(6,6)$ structure on $E$, which is reduced to SU$(3)\times$SU$(3)$ by the existence of the globally defined compatible spinors $\Psi_1$ and $\Psi_2$ \cite{Grana:2005sn}.

As discussed in more detail in \cite{Gualtieri:2007ng}, a pure spinor $\Psi$ is associated to an almost generalised complex structure, $\mathcal{J}$, whose integrability is equivalent to the following differential condition on the pure spinor 
\begin{equation}
\dd \Psi  =  V\cdot \Psi, 
\end{equation}
for some $V\in T\mathcal{M}\oplus T^\ast\mathcal{M}$. 

Thus the supersymmetry condition \eqref{ret} can be interpreted as the integrability of the generalised almost complex structure associated to the pure spinor $\Psi_2$, while the integrability of the one associated to $\Psi_1$ is obstructed by the RR-fluxes in \eqref{susy3}. 

This interpretation also allows to generalise the Hodge decomposition of tensors into holomorphic and anti-holomorphic components\footnote{Given the almost generalised complex structures action on differential forms \eqref{genstructact}.}. For instance, the generalised spinor bundle decomposes as 
\begin{align}
    \Lambda^\bullet T^\ast_M\otimes\mathbb{C}=&\bigoplus_{k=-3}^{3} U_k\\
    \Lambda^\bullet T^\ast_M\otimes\mathbb{C}=&\bigoplus_{k=-3}^{3} V_k,
\end{align} 
where $U_k$ and $V_k$ are subbundles of fixed eigenvalue of the generalised complex structures $\mathcal{J}_1$ and $\mathcal{J}_2$ respectively. The pair of almost generalised complex structures commute for SU$(3)\times$SU$(3)$ backgrounds, and one can decompose the space of complexified polyforms in terms of their common eigenspaces
\be  \Lambda^\bullet T^\ast_M\otimes\mathbb{C}=\bigoplus_{p,q} U_{p,q},
\label{Upq} \ee
where $U_{p,q}$ is the intersection of the $ip$-eigenspace of $\mathcal{J}_1$ and the $iq$-eigenspace of $\mathcal{J}_2$.
Given the non-integrability of the almost generalised complex structure $\mathcal{J}_1$ \eqref{susy3}, the exterior derivative decomposes into
\begin{align}
\dd_H\ \colon& \Gamma(U_k)\rightarrow \Gamma(U_{k-3})\oplus \Gamma(U_{k-1})\oplus\Gamma(U_{k+1})\oplus \Gamma(U_{k+3}),
\end{align}
while the integrability of the almost generalised complex structure $\mathcal{J}_2$ \eqref{ret} results in the following exterior derivative decomposition
\be 
\dd_H\colon \Gamma(V_k)\rightarrow \Gamma(V_{k-1})\oplus\Gamma(V_{k+1}).\label{extdersusy}
\ee

Each of the pure spinor equations \eqref{ret}, \eqref{second} and \eqref{susy3} has a clear interpretation as being a calibration condition for a certain type of D-brane, in the generalised sense. To make this statement precise, we briefly introduce the generalised geometry technology required to describe D-branes.

    In this formalism, a D-brane is characterised by two generalised ingredients, a generalised submanifold $(\Sigma,\mathcal{F})$ and a generalised calibration form $\omega$.
    
       A generalised submanifold is a pair $(\Sigma,\mathcal{F})$, with $\Sigma\subset\mathcal{M}$ a submanifold and $\mathcal{F}$ a two-form. The generalised submanifold is a generalised cycle if $\partial\Sigma=\emptyset$. A generalised submanifold wrapped by a D-brane respects \be \dd \mathcal{F}=H|_\Sigma,\label{Hint}\ee with $H|_\Sigma$ the pullback of the NSNS field-strength on the submanifold $\Sigma$.

        A generalised calibration form, which is a polyform $\omega$ of definite parity, can be associated to each D-brane. As shown in \cite{Martucci:2005ht}, one can construct  polyforms in terms of the pure spinors
 \begin{align}
 \label{caldef}
  \omega^{\nn{string}}=&\ e^{2A-\phi}\text{Im}\Psi_1\\
 \omega^\nn{DW}=&\ e^{3A-\phi}\Psi_2\\
     \omega^{\nn{sf}}=&\ e^{4A-\phi}\text{Re}\Psi_1,
 \end{align}
 which satisfy the properties of a generalised calibration. They first satisfy an algebraic condition corresponding to the minimisation of the D-brane energy
\begin{align}
    \mathcal{E}(\Pi,\mathcal{R})\geq\ (\omega|_\Pi\wedge e^\mathcal{R})(\Pi)\label{bound}
    \end{align}
for any point $p\subset\mathcal{M}$ and any generalised submanifold $(\Pi, \mathcal{R})$\footnote{Strictly speaking for any point $p\subset\mathcal{M}$ there must exist a generalised submanifold $(\Pi, \mathcal{R})$ such that the above bound is saturated.}. Here for $\rho$ a form, $\rho(\Pi)$ is the coefficient of the top form on $\Pi$ of $\rho|_\Pi$. The energy density corresponds to the following DBI contribution
 \be\mathcal{E}(\Pi,\mathcal{R})=e^{qA-\phi}\sqrt{\nn{det}(g|_\Pi+\mathcal{R})},\ee
 where $q$ is the number of external dimensions. Moreover, the differential conditions that must be respected by the above generalised calibration forms correspond to the supersymmetry conditions \eqref{ret}, \eqref{second}, and \eqref{susy3}
\begin{align}
  \dd_H(\omega^\nn{DW})=\ 0&\qquad\text{domain-wall BPSness}\\ 
  \dd_H(\omega^\nn{string})=\ 0&\qquad\text{D-string BPSness}\\
   \dd_H(\omega^\nn{sf})=\ e^{4A}\Tilde{F}&\qquad\text{gauge BPSness}.
\end{align}
%These are exactly the pure spinor equations required to preserve %\mathcal{N}=1$ supersymmetry, which is therefore equivalent to the calibration, or BPSness, of the would-be domain-wall, space-filling and string-like D-branes. 

\medskip

A generalised calibrated cycle is a generalised cycle saturating the calibration bound \eqref{bound}. A D-brane in an $\mathcal{N}=1$ background is supersymmetric, or BPS, if it wraps A generalised calibrated cycle.
This is why we refer to the $\mathcal{N}=1$ supersymmetry conditions as the domain-wall, D-string and gauge BPSness respectively. The above generalised calibration forms are associated to  string-like, domain-wall, and space-filling D-branes, which span respectively two, three and four non-compact dimensions.

\subsection{Reformulating the gauge BPSness}

In this paper, we consider non-supersymmetric solutions of type II supergravity, and we use first order differential equations on the pure spinors to describe them. These equations can be thought as modifications of the pure spinor equations \eqref{ret}, \eqref{second} and \eqref{susy3}, to which we add supersymmetry breaking terms, which is a standard procedure in the GCG literature \cite{Lust:2008zd,Legramandi:2019ulq,Menet:2023rnt}. Unlike the supersymmetric case, these modified pure spinor equations are not equivalent to the equations of motion, and
one has to also consider additional constraints to
have genuine $\mathcal{N}=0$ backgrounds.

   In this paper, we will only consider the case of $\mathcal{N}=0$ backgrounds which still respect the gauge BPSness \eqref{susy3}, therefore admitting stable (BPS) space-filling sources. We therefore consider generically
    \begin{align}
  \dd_H(\text{e}^{3A-\phi}\Psi_2)\neq&\ 0\label{retn0}\\ 
  \dd_H(\text{e}^{2A-\phi}\text{Im}\Psi_1)\neq&\ 0\label{dstringbpsn0}\\
   \dd_H(e^{4A-\phi}\text{Re}\Psi_1)=&\ e^{4A}\Tilde{\ast} F.\label{susy3bis}
\end{align}

It is more convenient to consider a different reformulation of the space-filling calibration condition \eqref{susy3bis}, which eliminate its implicit dependence on the metric coming from the Hodge operator, and express the RR field-strength itself in terms of the pure spinors.

    %These reformulations will prove to be useful to highlight the difference in the RR field-strength expressions, in the supersymmetric and non-supersymmetric case. They will also clarify the four-dimensional interpretation of \eqref{susy3bis}, in both the $\mathcal{N}=1$ and $\mathcal{N}=0$ cases.
    
        In the supersymmetric case, this reformulation was derived in \cite{Tomasiello:2007zq}:
\be F=-\dd_H^{\mathcal{J}_2}(e^{-\phi}\text{Re}\Psi_1)\label{Tomasiellosusy}\ee
with the introduction of the following differential operator
          \be \dd^\mathcal{J}=\text{e}^{-\frac{\pi}{2}\mathcal{J}\cdot}\dd\ \text{e}^{\frac{\pi}{2}\mathcal{J}\cdot}, \label{gendolbn0}\ee
where $\cdot$ denotes the action of the (almost) generalised complex structure on polyforms \eqref{genstructact} . When $\mathcal{J}_2$ is integrable, which is the case for $\mathcal{N}=1$ backgrounds, this differential reduces to \be \dd^\mathcal{J}=[\dd,\mathcal{J}\cdot].\ee
It is worth stressing that the derivation of \eqref{Tomasiellosusy}  relies on \eqref{second} being satisfied.

          Still in the supersymmetric case, a second reformulation of the gauge BPSness has been presented for instance in \cite{Koerber:2007xk}, and requires the introduction of the following complexification of the RR field-strength:
          \be G=F+i\dd_H(e^{-\phi}\nn{Re}\Psi_1).\label{G}\ee
          
          Decomposing \eqref{susy3} on the generalised Hodge diamond, and using again \eqref{second}, it can be shown that \eqref{susy3} is equivalent to 
          \begin{align}
              G|_{V_{-1}}=0\qquad G|_{V_{-3}}=0.\label{Gsusy}
          \end{align}
     
     The second equation is actually just
     \be F|_{V_{-3}}=0,\ee
     since the second term in $G$ is the derivative of a polyform of null charge under $\mathcal{J}_2$, which cannot have $V_{-3}$ components, from \eqref{extdersusy}. This reformulation is particularly useful to interpret \eqref{susy3} as an F-term condition in the four-dimensional theory, as discussed in \cite{Koerber:2007xk}.

        Let us now derive similar reformulations of the gauge BPSness \eqref{susy3bis}  in the non-supersymmetric case, where the domain-wall and D-string BPSness conditions are relaxed. 
        
        We generalise the derivation of \cite{Tomasiello:2007zq} in this case, and after some long but straightforward calculations we find
             \be F=-\dd_H^{\mathcal{J}_2}(e^{-\phi}\text{Re}\Psi_1)+e^{-2A}\mathcal{J}_1\cdot\dd_H^{\mathcal{J}_2}(e^{2A-\phi}\text{Im}\Psi_1).\label{fluxn0}\ee 
The second term in this expression explicitly shows the relationship between the additional RR flux components and the violation of the D-string BPSness \eqref{dstringbpsn0}. Moreover, the supersymmetry breaking terms that appear in \eqref{retn0}, which are in the 
$V_0$ bundle, prevent the generalised complex structure $\mathcal{J}_2$ from being integrable. As a consequence 
the exterior derivative property \eqref{extdersusy} is generalised to 
    \be \dd_H\colon\Gamma(V_k)\rightarrow \Gamma(V_{k-3})\oplus \Gamma(V_{k-1})\oplus\Gamma(V_{k+1})\oplus \Gamma(V_{k+3})\label{extdernonsusy},\ee
which also results in additional RR field-strength components compared to the supersymmetric case.

   From this reformulation of \eqref{susy3bis}, one can derive the non-supersymmetric analogue of \eqref{Gsusy}, by decomposing \eqref{fluxn0} on the generalised Hodge diamond.

        Looking first at the $V|_{-3}$ component, we now have
        \be F|_{V_{-3}}=i\dd_H(e^{-\phi}\text{Re}\Psi_1)|_{V_{-3}}.\ee

        The $V|_{-1}$ component is
        \be F|_{V_{-1}}=-i\dd_H(e^{-\phi}\text{Re}\Psi_1)|_{V_{-1}}+ie^{-2A}\mathcal{J}_1\cdot\dd_H(e^{2A-\phi}\text{Im}\Psi_1)|_{V_{-1}}.\ee
        These can be rearranged into
\begin{align}
G|_{V_{-1}}=&\ ie^{-2A}\mathcal{J}_1\cdot\dd_H(e^{2A-\phi}\text{Im}\Psi_1)|_{V_{-1}}\\
    G|_{V_{3}}=&\ 0,
\end{align}
which again is the most convenient form to make contact with the four-dimensional $\mathcal{N}=1$ effective theory. Indeed, an advantage of reformulating in this way the gauge BPSness for non-supersymmetric backgrounds is that it highlights the relationship between the complexified RR flux and the D-string BPSness violation \eqref{dstringbpsn0}. This is particularly interesting to discuss solutions dimensionally reducing to solutions of four-dimensional $\mathcal{N}=1$ supergravity, since the former enters in the corresponding on-shell superpotential, and the latter has been identified as the D-term contribution in four-dimensional $\mathcal{N}=1$ supergravity \cite{Koerber:2007xk}. We will address the four-dimensional $\mathcal{N}=1$ supergravity interpretation in more details in Section \ref{sec:n0}.

\section{4D structure and effective potential from pure spinors}\label{sec:effpot}

In this Section we briefly review $\mathcal{N} = 1$
Minkowksi solutions
of type II supergravity discussed in \cite{Koerber:2007xk}. More precisely, we discuss the interpretation of the pure spinor equations \eqref{ret}, \eqref{second} and \eqref{susy3} as the vanishing of some D-terms, F-terms and superpotential.
    
    We then write the most general four-dimensional `effective potential' from the ten-dimensional type II supergravity action, following \cite{Lust:2008zd}. Calling these scalar functions `effective potentials' is a bit misleading, since we write them as integral over the internal space, without choosing a specific truncation for the ten-dimensional modes and performing the actual dimensional reduction to write down a genuine scalar potential for the associated effective theories. However, in doing so we are able to interpret the different terms in the closed string sector of this effective potential as contributions from F-terms, D-terms and a superpotential. 

            In the next section, a similar interpretation in the non-supersymmetric case will motivate some constraints that the modified pure spinor equations should obey in order to be compatible with a four-dimensional $\mathcal{N}=1$ supergravity description with non-vanishing F-terms and D-terms.
        
\subsection{Four-dimensional $\mathcal{N}=1$ supergravity}\label{subsec:veffeomdwsb}

We briefly review here the work of \cite{Koerber:2007xk}, which provides ten-dimensional expressions for the four-dimensional $\mathcal{N}=1$ supergravity superpotential, D- and F-term conditions. The expression obtained in \cite{Koerber:2007xk} for the superpotential can be thought as the generalisation of the well-known Gukov-Vafa-Witten superpotential \cite{Gukov:1999ya} in the GCG formalism.

    We review these notions in order to then move on to the description of the effective potential for non-supersymmetric solutions of four-dimensional $\mathcal{N}=1$ supergravity, with non-vanishing D- and F-terms.
We introduce the following rescaled pure spinors
        \be t=e^{-\phi}\Psi_1\qquad\mathcal{Z}=e^{3A-\phi}\Psi_2,\ee
        as well as\begin{align}
    \mathcal{T}=\nn{Re}t-iC,\end{align}
    with $F=F^\nn{bg}+
\dd_H C$ and $F^\nn{bg}$ some fixed non-trivial background flux. Both $\mathcal{Z}$ and $\mathcal{T}$ are chiral fields of the associated four-dimensional $\mathcal{N}=1$ description. The polyform $\mathcal{T}$ is defined such that the complexified flux $G$ is its field-strength $G=i\dd_H\mathcal{T}$.

    We can now define the following superpotential and conformal Kähler potential densities
\be W=\pi(-1)^{|\mathcal{Z}|+1}\braket{\mathcal{Z},G}\qquad N=\frac{i\pi}{2}\braket{\mathcal{Z},\bar{\mathcal{Z}}}^{1/3}\braket{t,\bar{t}}^{2/3},\label{superpotkahlerdensity}
\ee
where $|\mathcal{Z}|$  is the degree (mod 2) of $\mathcal{Z}$,  which depend on both chiral fields $\mathcal{Z}$ and $\mathcal{T}$, and the associated superpotential and conformal Kähler potential
    \be \mathcal{W}=\int_{\mathcal{M}}W\qquad\mathcal{N}=\int_{\mathcal{M}}N.\label{superpotkahlerpot}\ee

Consider now a chiral field $X$ on which the superpotential and the conformal K\"ahler potential depend. Under a holomorphic variation $\delta X$ of $X$ we define 
        \be (\delta\mathcal{W})_X\coloneqq\delta_X\mathcal{W}-3(\delta_X\log \mathcal{N})\mathcal{W}. \ee
         We evaluate the variations associated to the two chiral fields at our disposal, $\mathcal{Z}$ and $\mathcal{T}$. First for $\mathcal{Z}$, we distinguish two contributions coming from two distinct holomorphic deformations of $\mathcal{Z}$
         \be \delta\mathcal{Z}\in (V_1\oplus V_3),\ee
         which we denote
          \be (\delta\mathcal{W})_{\mathcal{Z}_{(1)}}\quad\nn{and}\quad (\delta\mathcal{W})_{\mathcal{Z}_{(3)}}\ee
         respectively. They yield
         \begin{align}
     (\delta\mathcal{W})_{\mathcal{Z}_{(1)}}=&(-1)^{|\mathcal{Z}|+1}\pi\int_\mathcal{M}\left\langle\delta Z_{(1)},G|_{V_{-1}}\right\rangle\label{W1}\\
       (\delta\mathcal{W})_{\mathcal{Z}_{(3)}}=&-\pi\int_\mathcal{M}\left\langle\delta Z_{(3)},(-1)^{|\mathcal{Z}|} G|_{V_{-3}}+\frac{3i}{2}\frac{\mathcal{W}}{\mathcal{N}}e^{-4A}\bar{\mathcal{Z}}\right\rangle.\label{W2}
         \end{align}
    For $\mathcal{T}$, we find
        \be(\delta\mathcal{W})_\mathcal{T}=i  \pi\int_\mathcal{M}\left\langle\delta \mathcal{T},(-1)^{|\mathcal{Z}|}\dd_H \mathcal{Z}+3i\frac{\mathcal{W}}{\mathcal{N}}e^{2A}\nn{Im}t\right\rangle.\label{W3}\ee

Imposing the vanishing of the variations \eqref{W1}-\eqref{W3} reproduces the Anti de Sitter version of the supersymmetric pure spinor equations \eqref{ret} and \eqref{Gsusy}. If we also impose that the superpotential itself vanishes, they 
reduce to the corresponding equations in flat space
      \be \dd_H\mathcal{Z}=0\qquad G|_{V_{-1}}=0\qquad  G|_{V_{-3}}=0.\ee
        One can therefore interpret the two $\mathcal{N}=1$ supersymmetry conditions \eqref{ret} and \eqref{Gsusy} as F-term conditions and the vanishing of the superpotential
         \be \mathcal{W}=0\qquad (\delta\mathcal{W})_{\mathcal{Z}_{(1)}}=0\qquad (\delta\mathcal{W})_{\mathcal{Z}_{(3)}}=0\qquad (\delta\mathcal{W})_{\mathcal{T}}=0.\ee

            On another note, the parametrisation in terms of the chiral fields $\mathcal{T}$ and $\mathcal{Z}$ has some redundancy, due to the the RR gauge transformations $\delta_\lambda C=\dd_H\lambda$ resulting in $\delta_\lambda\mathcal{T} =-i\dd_H\lambda$. These symmetries are gauged in the effective theory, and their associated D-terms have been worked out in \cite{Koerber:2007xk}, yielding
            \be \mathcal{D}(\lambda)=2\pi\int_{\mathcal{M}}\braket{\lambda,\mathcal{D}}\ee
            with \be \mathcal{D}=\dd_H(e^{2A}\nn{Im}t),\label{Dterms}\ee
            the D-term density.
            The last $\mathcal{N}=1$ supersymmetry condition \eqref{second} can thus be interpreted as the vanishing of these D-terms \be \mathcal{D}=0,\ee
            completing the four-dimensional picture.

            One can also write down the corresponding covariant derivatives of the superpotential density. For a chiral field X
            \be\mathcal{D}_XW\equiv\partial_XW-3(\partial_X\log N)W.\ee
            In the case of $\mathcal{N}=1$ Minkowski solutions, they are\footnote{We deliberately give the expressions with vanishing superpotential in order to make contact with the on-shell $\mathcal{N}=1$ Minkowski effective potential discussed in the next subsection.}
              \begin{align}
        \mathcal{D}_{\mathcal{Z}_{(1)}}W=&\pi(-1)^{|\mathcal{Z}|+1} G|_{V_{-1}}\label{DWG1}\\
        \mathcal{D}_{\mathcal{Z}_{(3)}}W=& \pi(-1)^{|\mathcal{Z}|+1} G|_{V_{-3}}\label{DWG3}\\
        \mathcal{D}_{\mathcal{T}}W=&i\pi(-1)^{|\mathcal{Z}|} \dd_H\mathcal{Z}.\label{DWdZ}
         \end{align}
        \subsection{The type II effective potential }\label{subsec:veff}

In this subsection, we follow \cite{Lust:2008zd} and give the four-dimensional effective action for backgrounds with ten-dimensional space-time of the form $X_4 \times \mathcal{M}$ with the metric
\be \dd s^2_{10}=e^{2A(y)}g_{\mu\nu}\dd x^\mu\dd x^\nu+g_{mn}\dd y^m\dd y^n,\ee
where $g_{\mu\nu}$ is for now a general four-dimensional metric depending only on the external coordinates, and all the other fields depend only on the internal coordinates. 

    The effective four-dimensional action is\footnote{We use the convention $2\pi\sqrt{\alpha'}=1$, so that all D-brane
tensions are equal. We are also neglecting anomalous curvature-like corrections to the sources
contribution: they can be easily added without affecting the results of the
discussion. Finally we are also omitting the internal field kinetic terms, since they are taken to be constant along the external directions.}

    \be S_{\text{eff}}=\int_{X_4} \dd^4x\sqrt{-g_4}\left(\frac{1}{2}\mathcal{N}R_4-2\pi\mathcal{V}_{\text{eff}}\right),\label{seff}\ee
    where $R_4$ is the four-dimensional scalar curvature, and
    \begin{align}
\mathcal{V}_{\nn{eff}}=&\int_{\mathcal{M}}\nn{vol}_6 e^{4A}\{e^{-2\phi}[-\mathcal{R}+\frac{1}{2}H^2-4(\dd\phi)^2+8\nabla^2A+20(\dd A)^2]-\frac{1}{2}\Tilde{F}^2\}\nonumber\\
        &+\sum_{i \in \nn{loc. sources}}\tau_i\left(\int_{\Sigma_i}e^{4A-\phi}\sqrt{\nn{det}(g|_{\Sigma_i}+\mathcal{F}_i)} -\int_{\Sigma_i}C^\nn{el}|_{\Sigma_i}\wedge e^{\mathcal{F}_i}\right)\label{vefff}
    \end{align} 
   is the type II effective potential density, with $\mathcal{R}$ the six-dimensional scalar curvature. Its first line corresponds to the closed string sector, while the second line is the localised sources contributions. For the O-planes, we set $\mathcal{F}=0$, and we have $\tau_{\nn{D}_p}=1,\ \tau_{O_q}=-2^{q-5}$.
   The sources couple to the RR potentials defined by $\dd_HC^{\nn{el}}=e^{4A}\Tilde{F}$.

 The variations of the four dimensional action \eqref{seff} exactly reproduce the ten-dimensional equations of motion, as argued in \cite{Lust:2008zd}. They are given in Appendix \ref{sec:AppA}. Moreover, from the variation with respect to $g_{\mu\nu}$, one gets that the external space is Einstein, with \be R_4=8\pi\mathcal{V}_\text{eff}/\mathcal{N},\ee as expected. From the ten-dimensional perspective this is equivalent to the internal space integral of the external ten-dimensional Einstein equation's trace. As we consider Minkowski backgrounds, we focus on the cases where the effective potential vanishes at the solutions.

        Finally, the RR equations of motion reproduce the usual Bianchi identities
        \be \dd_HF=-j_\text{tot}=-\sum_{i}\tau_ij_i,\label{bianchi}\ee
        where, as described in \cite{Lust:2008zd} for instance, the $j_i$ are the generalised currents for the localised sources.

        From now on we will use the rewriting of the effective potential in terms of the pure spinors, derived in \cite{Lust:2008zd}:
        \begin{align}
        \mathcal{V}_\text{eff}=&
         \frac{1}{2}\int_{\mathcal{M}}\text{vol}_6  \Big[|\dd_H(e^{2A-\phi}\nn{Im}\Psi_1)|^2
         + e^{-2A}  |\dd_H(e^{3A-\phi}\Psi_2)|^2\Big]\nonumber\\
         &+\frac{1}{2}\int_{\mathcal{M}}\text{vol}_6 \ e^{4A}|\Tilde{\ast} F- e^{-4A}\dd_H(e^{4A-\phi}\text{Re}\Psi_1)|^2\nonumber\\
    &-\frac{1}{4}\int_{\mathcal{M}}e^{-2A}\left(\frac{|\braket{\Psi_1,\dd_H(\text{e}^{3A-\phi}\Psi_2)}|^2}{\text{vol}_6}+\frac{|\braket{\bar{\Psi}_1,\dd_H(\text{e}^{3A-\phi}\Psi_2)}|^2}{\text{vol}_6}\right)\nonumber\\
    &-4\int_{\mathcal{M}}\text{vol}_6 e^{4A-2\phi}[(u^1_R)^2+(u^2_R)^2]\nonumber\\
     &+\sum_{i\subset\text{D-branes}}\tau_i \int_{\mathcal{M}}e^{4A-\phi}(\text{vol}_6 \ \rho_i^\text{loc}-\braket{\text{Re}\Psi_1, j_i})\nonumber\\
&+\int_{\mathcal{M}}\braket{e^{4A-\phi}\text{Re}\Psi_1-C^\text{el},\dd_H F+j_\text{tot}}.\label{effpot}
    \end{align}
        The square of a polyform is defined in Appendix \ref{sec:AppA}, and we have 
          \be u^{1,2}_R= u^{1,2}_{Rm}\dd y^m\equiv(u^{1,2}_m+u^{\ast 1,2}_m)\dd y^m,\label{uR}\ee
with 
    \begin{align}
u^1_m=&\frac{i\braket{\gamma_m\bar{\Psi}_1,\dd_H(\text{e}^{2A-\phi}\text{Im}\Psi_1)}}{e^{2A-\phi}\braket{\Psi_1,\bar{\Psi}_1}}+\frac{\braket{\gamma_m\bar{\Psi}_2,\dd_H(\text{e}^{3A-\phi}\Psi_2)}}{2e^{3A-\phi}\braket{\Psi_2,\bar{\Psi}_2}}\label{u1}\\
u^2_m=&\frac{i(-1)^{|\Psi_2|}\braket{\Psi_1\gamma_m,\dd_H(\text{e}^{2A-\phi}\text{Im}\Psi_1)}}{e^{2A-\phi}\braket{\Psi_1,\bar{\Psi}_1}}+\frac{(-1)^{|\Psi_1|}\braket{\bar{\Psi}_2\gamma_m,\dd_H(\text{e}^{3A-\phi}\Psi_2)}}{2e^{3A-\phi}\braket{\Psi_2,\bar{\Psi}_2}}.\label{u2}
    \end{align}
    The gamma matrix conventions are given in Appendix \ref{sec:AppA}. We also introduced here the Born-Infeld density $\rho^\nn{loc}_i$ associated with a source wrapping a generalised submanifold $(\Sigma_i,\mathcal{F}_i)$
    \be \rho_i^\nn{loc}=\frac{\sqrt{\nn{det}(g|_{\Sigma_i}+\mathcal{F}_i)}}{\sqrt{\nn{det}g}}\delta(\Sigma_i).\ee

    It is useful to rewrite the algebraic inequality \eqref{bound} in terms of $\rho^\nn{loc}$:
     \be \rho^\nn{loc}_i\geq\frac{\braket{\nn{Re}\Psi_1,j_i}}{\text{vol}_6},\label{boundrho}\ee
     where the division by the volume form means that we remove the vol$_6$ factor in the numerator.

        Therefore, if one considers the effective potential of a (non-)supersymmetric background with calibrated sources, the above inequality is saturated and the fifth line of \eqref{effpot} vanishes. Similarly, the last line of \eqref{effpot} will vanish for solutions of ten-dimensional supergravity satisfying the Bianchi identities \eqref{bianchi}.  

In the case of an $\mathcal{N}=1$ Minkowski background, the closed-string sector of the on-shell effective potential can be rewritten as
\begin{align}
        \mathcal{V}_\text{eff}=&
         \frac{1}{2}\int_{\mathcal{M}}  e^{4A}(|G_{-1}|^2+|G_{-3}|^2)
         +  e^{-2A} |\dd_H(\mathcal{Z})|^2-\frac{e^{-2A+2\phi}}{8\nn{vol}_6}|\braket{\mathcal{Z},G}|^2\nonumber\\
          &+\frac{1}{2}\int_{\mathcal{M}}|\dd_H(e^{2A-\phi}\text{Im} \Psi_1)|^2.\label{n1Veff}
    \end{align}
Though this is simply the
$\mathcal{N}=1$ on-shell
effective potential, this is the natural formulation to interpret each term as the vanishing of some D-terms, F-terms and superpotential.

    The first three terms in the first line of \eqref{n1Veff} can be identified with the covariant derivatives of the superpotential density \eqref{DWG1}, \eqref{DWG3} and \eqref{DWdZ} respectively, while the last term in the first line of \eqref{n1Veff} can be identified with the superpotential density \eqref{superpotkahlerdensity}. Finally, the last line of \eqref{n1Veff} can be identified with the D-terms \eqref{Dterms}.

    We do not intend to make a rigorous identification
with the usual four-dimensional $\mathcal{N}=1$ scalar potential here, we simply stress that each term in the closed-string sector of the effective potential of ten-dimensional $\mathcal{N}=1$ type II supergravity Minkowski solutions fits into the four-dimensional $\mathcal{N}=1$ supergravity description.
    
   % Inspired by this fact, we then want to see under which conditions one can carry a similar interpretation of the on-shell effective potential for non-supersymmetric ten-dimensional type II supergravity Minkowski backgrounds, which would then dimensionally reduce to four-dimensional $\mathcal{N}=1$ supergravity solutions with non-vanishing D-, F-terms, and superpotential.%
             
    \section{D-terms in Generalised Complex Geometry}\label{sec:n0}

    In this Section we investigate the on-shell effective potential of non-supersymmetric solutions of type II supergravity with external Minkowski space. We identify some conditions that the modified pure spinor equations must satisfy in order for the non-supersymmetric solutions to have a clear interpretation in terms of four-dimensional $\mathcal{N}=1$ supergravity.  

We also derive the equations of motion, in the language of pure spinors, associated to these backgrounds, with and without D-terms. 
   \subsection{Effective potential and F-term conditions}
   Let us recall that we focus on non-supersymmetric solutions having only space-filling sources, in order to preserve the Poincar\'e symmetry of the external space, and that we consider only BPS sources. We also consider that our backgrounds satisfy the Bianchi identities \eqref{bianchi}.
   
   In this case, the type II effective potential is
       \begin{align}
        \mathcal{V}_\text{eff}=&
         \frac{1}{2}\int_{\mathcal{M}}\text{vol}_6  \Big[|\dd_H(e^{2A-\phi}\nn{Im}\Psi_1)|^2
         + e^{-2A}  |\dd_H(e^{3A-\phi}\Psi_2)|^2\Big]\nonumber\\
         &-\frac{1}{4}\int_{\mathcal{M}}e^{-2A}\left(\frac{|\braket{\Psi_1,\dd_H(\text{e}^{3A-\phi}\Psi_2)}|^2}{\text{vol}_6}+\frac{|\braket{\bar{\Psi}_1,\dd_H(\text{e}^{3A-\phi}\Psi_2)}|^2}{\text{vol}_6}\right)\nonumber\\
         &+\frac{1}{2}\int_{\mathcal{M}}\text{vol}_6 \ e^{4A}|\Tilde{\ast} F- e^{-4A}\dd_H(e^{4A-\phi}\text{Re}\Psi_1)|^2\nonumber\\
    &-4\int_{\mathcal{M}}\text{vol}_6 e^{4A-2\phi}[(u^1_R)^2+(u^2_R)^2]\label{effpotn0}.
    \end{align}
    For non-supersymmetric Minkowski solutions, this potential still vanishes, but each term does not necessarily vanish identically.

    We now want to restrict ourselves to the study of a subclass of backgrounds dimensionally reducing to a solution of $\mathcal{N}=1$ four-dimensional supergravity, in the sense that their scalar potential can be written as F-terms, D-terms, and superpotential contributions. From the ten-dimensional perspective, this means that we focus on backgrounds where each terms in the on-shell scalar potential \eqref{effpotn0} has an interpretation in terms of the aforementioned contributions.

    At this point the most pressing question to address is therefore the superpotential expression for such backgrounds.

    As mentioned above, we consider backgrounds where supersymmetry is broken in a controlled way, as a perturbation around a certain supersymmetric backgrounds. For instance, we could think of the right-hand side contributions of the pure spinor equations \eqref{retn0} and \eqref{dstringbpsn0} as controlled by some supersymmetry breaking parameters, whose vanishing would restore supersymmetry. The question now becomes: how is the superpotential in  \eqref{superpotkahlerpot} affected by switching on right-hand side contributions in \eqref{retn0} and \eqref{dstringbpsn0}.

    In the supersymmetric case, the complexification of the RR potentials entering the superpotential is suggested by the coupling of a BPS space-filling D-brane to magnetic background fields \cite{Martucci:2006ij,Koerber:2007xk}. Indeed, the action of a space-filling D-brane wrapping a generalised cycle $(\Sigma,\mathcal{F})$ can be written as\footnote{The standard space-filling D-brane action is here Wick rotated to Euclidean space.}
    \be S_{\text{D-brane}}=\int_\Sigma e^{4A-\phi}\sqrt{\text{det}(g|_\Sigma+\mathcal{F})}\text{vol}_\Sigma-ie^{4A}C|_\Sigma\wedge e^\mathcal{F}\, ,\ee
    with vol$_\Sigma$ the volume form on the cycle $\Sigma$. The calibration of the space-filling D-brane imposes that
    \be e^{4A-\phi}\sqrt{\text{det}(g|_\Sigma+\mathcal{F})}\text{vol}_\Sigma=e^{4A-\phi}\text{Re}\Psi_1\wedge e^\mathcal{F}\, ,\label{calibsf}\ee
so the resulting action is
 \be S_{\text{D-brane}}=\int_\Sigma e^{4A}(e^{-\phi}\text{Re}\Psi_1-iC)\wedge e^\mathcal{F}=\int_\Sigma e^{4A}\mathcal{T}\wedge e^\mathcal{F}\, ,\label{actionsfdb}\ee
    putting in evidence $\mathcal{T}$ as the natural complexification of the RR potentials.

    We now consider the breaking of supersymmetry through switching on right-hand side contributions in \eqref{retn0} and \eqref{dstringbpsn0}. Crucially, the complexification of the RR flux entering the superpotential is unaltered. Indeed, this is because the supersymmetry breaking perturbations are so that the calibration of the space-filling D-branes \eqref{calibsf} is preserved, and thus the expression of the BPS space-filling D-brane action \eqref{actionsfdb} holds. The superpotential expression thus remains as in \eqref{superpotkahlerpot}, although it must be non-vanishing on-shell for non-supersymmetric Minkowski backgrounds, as it is the only negative contribution to the vanishing effective potential. The associated F-terms might not be vanishing either.
    
    Non-supersymmetric backgrounds with BPS space-filling sources have been studied in the GCG literature, and the contributions to the scalar potential of these vacua have indeed been interpreted as F-terms and a superpotential contribution with the superpotential \eqref{superpotkahlerpot} \cite{Lust:2008zd}. It is also interesting to mention that backgrounds respecting the algebraic calibration condition \eqref{calibsf} but not the differential one \eqref{susy3bis} are successfully described using the superpotential \eqref{superpotkahlerpot} in \cite{Grana:2020hyu}.
    
    Generically we have no way to control the superpotential expression for non-supersymmetric backgrounds with non-BPS sources dimensionally reducing to solutions of four-dimensional $\mathcal{N}=1$ supergravity. We thus don't further address this case.

    Coming back to the scalar potential \eqref{effpotn0}, the BPSness of the space-filling sources \eqref{susy3bis} implies that its third line must vanish on-shell. In the supersymmetric case, this corresponds to the F-term conditions resulting from the vanishing of the variations $(\delta\mathcal{W})_{\mathcal{Z}_{(1)}}$ and $(\delta\mathcal{W})_{\mathcal{Z}_{(-3)}}$.We investigate these conditions when the supersymmetry breaking contributions are switched on. Given that the BPSness of the space-filling sources is preserved on-shell, the third line of the effective potential \eqref{effpotn0} still vanishes and thus the F-term conditions from $(\delta\mathcal{W})_{\mathcal{Z}_{(1)}}$ and $(\delta\mathcal{W})_{\mathcal{Z}_{(-3)}}$ should keep on vanishing.

    Recall from the previous Subsection that the variation of the superpotential with respect to $\mathcal{Z}_{(1)}$ reads
    \be  (\delta\mathcal{W})_{\mathcal{Z}_{(1)}}=(-1)^{|\mathcal{Z}|+1}\pi\int_\mathcal{M}\left\langle\delta Z_{(1)},G|_{V_{-1}}\right\rangle.\ee
    The resulting F-term condition is therefore
    \be G|_{V_{-1}}=0.\label{Ftermn0}\ee
However, in the case of generic domain-wall and D-string BPSness violation,  \eqref{susy3bis} is equivalent to
    \begin{align}
G|_{V_{-1}}=&\ ie^{-2A}\mathcal{J}_1\cdot\dd_H(e^{2A-\phi}\text{Im}\Psi_1)|_{V_{-1}}\label{Gm1}\\
    G|_{V_{3}}=&\ 0.
\end{align}
The F-term condition \eqref{Ftermn0} thus results in
\be \mathcal{J}_1\cdot\dd_H(e^{2A-\phi}\text{Im}\Psi_1)|_{V_{-1}}=0.\ee
This is equivalent to
\be \dd_H(e^{2A-\phi}\text{Im}\Psi_1)|_{U_{2,-1}}=\dd_H(e^{2A-\phi}\text{Im}\Psi_1)|_{U_{-2,-1}}=0 ,\ee
where the sub-spaces $U_{p,q}$ are defined in \eqref{Upq}.
Given that the polyform $\dd_H(e^{2A-\phi}\text{Im}\Psi_1)$ is real, it implies
\be \dd_H(e^{2A-\phi}\text{Im}\Psi_1)\subset U_0.\label{condDterm}\ee
Backgrounds which don't respect \eqref{condDterm} fall off of the class of solutions dimensionally reducing to solutions of four-dimensional $\mathcal{N}=1$ supergravity with the superpotential \eqref{superpotkahlerpot}. The effective theories associated to such solutions could be described as solutions of four-dimensional $\mathcal{N}=1$ supergravity with a different superpotential\footnote{And therefore not admitting BPS space-filling D-branes.}, or could be described with a fake superpotential. Alternatively, it might not even be sensible to talk about effective theories associated to these ten-dimensional backgrounds, or it could be that their effective theories are non-supersymmetric with the field content of $\mathcal{N} =1$ or $\mathcal{N} =2$ supergravity with additional (massive) multiplets, or non-supersymmetric solutions of four-dimensional supergravities with higher supersymmetry. We don't address these possibilities further and for the rest of this paper we focus on the backgrounds respecting \eqref{condDterm}.

Interestingly, imposing the condition \eqref{condDterm} makes the last line of the potential \eqref{effpotn0} vanish\footnote{This can be seen by imposing \eqref{condDterm} on the SU$(3)\times$SU$(3)$ decomposition of the pure spinor equations given in Appendix \ref{sec:AppC}.}
\be u^1_R=u^2_R=0\label{ucond}.\ee
This last condition turns out to be crucial for the associated four-dimensional $\mathcal{N}=1$ supergravity, given that the last line of the potential \eqref{effpotn0} is negative semi-definite and is not a superpotential contribution, so it must indeed vanish on-shell, which is guaranteed by \eqref{ucond}.

The modes set to zero by \eqref{condDterm} belong to vector representations under the SU$(3)\times$SU$(3)$ structure, just like the massive spin $\frac{3}{2}$-multiplet degrees of freedom that appear when reducing the ten-dimensional theory to four-dimensions \cite{Grana:2009im,Grana:2011nb}, which are seen as non-physical degrees of freedom in the four-dimensional $\mathcal{N}=1$ supergravity and should be gauged away. It is then reasonable to interpret our condition \eqref{condDterm} as the requirement to keep only the four-dimensional $\mathcal{N}=1$ multiplets in the low-energy effective theories.

    On another note, the violation of the domain-wall BPSness condition \eqref{retn0} results in a non vanishing variation of the superpotential \eqref{W3}. The corresponding contribution to the superpotential (the second term of the first line of \eqref{effpotn0}) is therefore naturally interpreted as an F-term.

    The condition \eqref{condDterm} ensures the vanishing of the F-term corresponding to the calibration of space-filling D-branes, but of course it does not guarantee that non-supersymmetric ten-dimensional solutions respecting this condition will have a four-dimensional $\mathcal{N}=1$ low-energy effective theory.

    The type II supergravity solutions actually reducing to four-dimensional $\mathcal{N}=1$ models with vanishing D-terms automatically obey \eqref{condDterm}, since then $\dd_H(e^{2A-\phi}\text{Im}\Psi_1)=0$. This is the case for the GKP-like solutions of \cite{Giddings:2001yu,Lust:2008zd} for example, but more generally it is true for backgrounds which reduce to no-scale models with non-vanishing F-terms and superpotential. In the GCG literature, the non-supersymmetric type II supergravity solutions with a supersymmetry breaking term violating the D-string BPSness \eqref{dstringbpsn0} do not respect the condition \eqref{condDterm}, and therefore do not reduce to solutions of four-dimensional $\mathcal{N}=1$ supergravity \cite{Legramandi:2019ulq, Menet:2023rnt}. In the next Section, we will investigate the constraint \eqref{condDterm} in a more concrete setting.
    
   \subsection{Equations of motion}
   In this subsection we derive the equations of motion in the language pure spinors for the class of backgrounds discussed above, admitting BPS space-filling sources and respecting the condition \eqref{condDterm} as well as the Bianchi identities. We do so by requiring the vanishing of the variations of the effective potential with respect to the internal fields, which is equivalent to the ten-dimensional equations of motion given in Appendix \ref{sec:AppA}, (see \cite{Lust:2008zd}). 

  We consider the following potential\footnote{The other terms in the effective potential give trivial contributions to the equations of motion since they are quadratic in quantities vanishing for the considered backgrounds.}
   \begin{align}
        \mathcal{V}_\text{eff}=
         &\ \frac{1}{2}\int_{\mathcal{M}} e^{-2A}    \braket{\Tilde{\ast}_6 [\dd_H(\text{e}^{3A-\phi}\Psi_2)],\dd_H(\text{e}^{3A-\phi}\bar{\Psi}_2)}\nonumber\\
         &
         +\frac{1}{2}\int_{\mathcal{M}}  \braket{\Tilde{\ast}_6 [\dd_H(\text{e}^{2A-\phi}\text{Im}\Psi_1)],\dd_H(\text{e}^{2A-\phi}\text{Im}\Psi_1)}\nonumber\\
    &-\frac{1}{4}\int_{\mathcal{M}}e^{-2A}\left(\frac{|\braket{\Psi_1,\dd_H(\text{e}^{3A-\phi}\Psi_2)}|^2}{\text{vol}_6}+\frac{|\braket{\bar{\Psi}_1,\dd_H(\text{e}^{3A-\phi}\Psi_2)}|^2}{\text{vol}_6}\right).
    \label{FFDpot}
    \end{align}
    We introduce the polyforms
    \begin{align}
    \label{Bth}
\Theta=&e^{-2A}\Tilde{\ast}_6\dd_H(e^{3A-\phi}\Psi_2)+2i(-1)^{|\Psi_1|}e^{A-\phi}(\bar{t}_1\Psi_1+\bar{t}_2\bar{\Psi}_1)\\
\label{Bch}\Xi=&\Tilde{\ast}_6\dd_H(\text{e}^{2A-\phi}\text{Im}\Psi_1),
\end{align}
    with
    \be t_1=2(-1)^{|\Psi_1|}\frac{\braket{\dd_H(e^{3A-\phi}\Psi_2),\Psi_1}}{\braket{\Psi_1,\bar{\Psi}_1}}\qquad t_2=2(-1)^{|\Psi_1|}\frac{\braket{\dd_H(e^{3A-\phi}\Psi_2),\bar{\Psi}_1}}{\braket{\Psi_1,\bar{\Psi}_1}}.
    \ee
    The decompositions of $\Theta$ and $\Xi$ on the SU$(3)\times$SU$(3)$ structure are given  in Appendix \ref{sec:AppC}.

Then, varying the potential \eqref{FFDpot} with respect to the dilaton, we find the following dilaton equation of motion
     \begin{align}
       \nn{Re}\{\braket{e^{3A-\phi}\Psi_2,\dd_H\Theta}\}+\braket{e^{2A-\phi}\nn{Im}\Psi_1,\dd_H\Xi}=0.
    \end{align}
Note that, in our case,  the  solutions of the dilaton equation
have an identically vanishing effective potential.
    The B-field equation of motion is
     \be \dd\Big[\nn{Re}\{\braket{e^{3A-\phi}\Psi_2,\Theta}_3\}+\braket{e^{2A-\phi}\nn{Im}\Psi_1,\Xi}_3\Big]=0.\label{Bfieldeom}\ee
    
    We derive the internal component of the Einstein equation by varying the effective potential with respect to the internal metric. Given that the Hodge operator and the pure spinors depend implicitly on the metric, we use the following rules
    \begin{align}
                \delta\sqrt{\nn{det}g}=&-\frac{1}{2}\delta g^{mn}g_{mn}\sqrt{\nn{det}g}\\
    \delta\braket{\Tilde{\ast}_6 \chi_1,\chi_2}=&\ \delta g^{mn}\big[\braket{\Tilde{\ast}_6 \iota_m\chi_1,\iota_n\chi_2}-\frac{1}{2}g_{mn}\braket{\Tilde{\ast}_6 \chi_1,\chi_2}\big]\\
    \delta\Psi_i=&-\frac{1}{2}\delta g^{mn}g_{k(m}\dd y^k\wedge\iota_{n)}\Psi_i\qquad i=1,2.
            \end{align}
    We find the following internal Einstein equations
    \begin{align}
        0=& \nn{Re}\Big\{\braket{g_{k(m}\dd y^k\wedge\iota_{n)}(e^{3A-\phi}\Psi_2),\dd_H\Theta}-\braket{g_{k(m}\dd y^k\wedge\iota_{n)}\dd_H(e^{3A-\phi}\Psi_2),\Theta}\Big\}\nonumber\\
         &+\braket{g_{k(m}\dd y^k\wedge\iota_{n)}(e^{2A-\phi}\nn{Im}\Psi_1),\dd_H\Xi}-\braket{g_{k(m}\dd y^k\wedge\iota_{n)}\dd_H(e^{2A-\phi}\nn{Im}\Psi_1),\Xi}.
    \end{align}
    These are equations of motion of backgrounds compatible with a four-dimensional supergravity solutions with non vanishing F-terms and D-terms.

    A subclass of these backgrounds are those  compatible with a four-dimensional supergravity solutions with non vanishing F-terms and vanishing D-terms. In this class we have
    \be \dd_H(e^{2A-\phi}\text{Im}\Psi_1)=0,\ee
    so obviously $\Xi=0$. 
    
    However, the last equation of motion, the external component of the modified Einstein equation, must be discussed separately for the cases with and without D-terms.

   The external component of the modified Einstein equation is equivalent to the vanishing of the following variation of the effective potential \be 
\frac{\delta \mathcal{V}_\nn{eff}}{\delta A}+2 \frac{\delta \mathcal{V}_\nn{eff}}{\delta \phi}=0,\label{modeinstein2}\ee and it is identically satisfied for backgrounds with calibrated sources, preserving the D-string BPSness (i.e. without D-terms) and satisfying the Bianchi identities, as shown in \cite{Lust:2008zd}. 

For backgrounds with non-vanishing D-terms, thus violating the D-string BPSness, we simply reduce 
the ten-dimensional equation \eqref{modeinstein} on our warped configurations 
\be 
\nabla^m(e^{-2\phi}\nabla_me^{4A})=e^{4A}\Tilde{F}\cdot\Tilde{F}+e^{4A-\phi}\sum_{i \in \nn{loc.~ sources}}\tau_i\rho_i^\nn{loc},
\ee
and rewrite it in terms of pure spinors as
\begin{align} 
\label{extmodE}
-\dd(e^{-2\phi}\ast_6\dd e^{4A})=&\braket{\Tilde{\ast}_6 \Tilde{F},e^{4A}\Tilde{F}}-\braket{\dd_H\Tilde{\ast}_6 \Tilde{F},e^{4A-\phi}\nn{Re}\Psi_1} \nonumber \\
&+e^{4A-\phi}\sum_{i \in \nn{loc. sources}}\tau_i\Big[\rho_i^\nn{loc}\nn{vol}_6-\braket{\nn{Re}\Psi_1,j_i}\Big]
\end{align}
by using the Bianchi identity \eqref{bianchi} together with the RR-field-strength self-duality \eqref{fselfdual}. For backgrounds admitting only calibrated sources, the second line in \eqref{extmodE} vanishes, and we are left with  the external components of the modified Einstein equation
\begin{align} 
-\dd(e^{-2\phi}\ast_6\dd e^{4A})=&\braket{\Tilde{\ast}_6 \Tilde{F},e^{4A}\Tilde{F}}-\braket{\dd_H\Tilde{\ast}_6 \Tilde{F},e^{4A-\phi}\nn{Re}\Psi_1}.
\end{align}
These equations of motion are drastically simpler than the ones one would obtain by varying the effective potential \eqref{effpotn0}.  Considering the case where $\Xi=0$, the complete set of equations of motion
\begin{gather}
       \nn{Re}\{\braket{e^{3A-\phi}\Psi_2,\dd_H\Theta}\}=0\\
       \dd\Big[\nn{Re}\{\braket{e^{3A-\phi}\Psi_2,\Theta}_3\}\Big]=0\\
      \nn{Re}\Big\{\braket{g_{k(m}\dd y^k\wedge\iota_{n)}(e^{3A-\phi}\Psi_2),\dd_H\Theta}-\braket{g_{k(m}\dd y^k\wedge\iota_{n)}\dd_H(e^{3A-\phi}\Psi_2),\Theta}\Big\}=0
            \end{gather}
    is simple enough to hope to solve them for new ten-dimensional type II flux vacua with BPS space-filling sources, which would dimensionally reduce to solutions of four-dimensional $\mathcal{N}=1$ supergravity with non-vanishing F-terms, like no-scale models for example.

       \section{The example of type IIB SU$(3)$-backgrounds with BPS O5-planes}\label{sec:example}
In this Section, we investigate again the condition \eqref{condDterm}
we have to impose in order to interpret our solutions as dimensionally reducing to $\mathcal{N}=1$ four-dimensional supergravity. In particular, we want to determine how restrictive the condition \eqref{condDterm} can be. Concretely, we would like to consider the possibility of having non-zero D-terms $\dd_H(e^{2A-\phi}\text{Im}\Psi_1)\neq 0$, and see how constrained their expression is from  requiring $\dd_H(e^{2A-\phi}\text{Im}\Psi_1)\in U_0$.

We focus on  type IIB warped backgrounds with a four-dimensional Minkowski external space, admitting calibrated parallel space-filling O5-planes, and possibly D5-branes, and we restrict to internal manifolds with SU$(3)$ structure.  Introducing a local unwarped vielbein $\{\Tilde{e}^a\}$, we choose the directions $\Tilde{e}^1$ and $\Tilde{e}^4$ to be tangent to the unique two-cycle wrapped by the sources. Our metric ansatz is thus \begin{align}
    \dd s^2=&\text{e}^{2A}\dd s^2_{\mathbb{R}_{1,3}}+  \dd s^2_{\mathcal{M}}\\
    \dd s^2_{\mathcal{M}}=&\text{e}^{2A}[(\Tilde{e}^1)^2+(\Tilde{e}^4)^2]+\text{e}^{-2A}\sum_{j=2,3,5,6}(\Tilde{e}^j)^2.
    \end{align}

For SU$(3)$-structure manifolds, the pure spinors \eqref{puresp1} and \eqref{puresp2} reduce to 
\be \Psi_1=e^{i\theta}e^{iJ}\qquad\Psi_2=e^{-i\theta}\Omega , \ee
where $\theta$ is the relative phase between the two parallel internal spinors  $\eta_1=ie^{i\theta}\eta_2$, the 
 Kähler form $J$ and the $(3,0)$ form $\Omega$ take the form
\begin{align}
    J=&-(e^{2A}\Tilde{e}^1\wedge \Tilde{e}^4+e^{-2A}\Tilde{e}^2\wedge \Tilde{e}^5+e^{-2A}\Tilde{e}^3\wedge \Tilde{e}^6)\\
    \Omega=&e^{-A}(\Tilde{e}^1+i\Tilde{e}^4)\wedge(\Tilde{e}^2+i\Tilde{e}^5)\wedge(\Tilde{e}^3+i\Tilde{e}^6).
\end{align} 
As discussed in \cite{Camara:2007cz}, the orientifold projection sets
\be \theta=-\frac{\pi}{2}.\ee

Notice that,  combining the dilaton equation of motion with the appropriately traced external components of the Einstein equations, as done in \cite{Blaback:2010sj}, one can show that\footnote{Note that both the Bianchi identities and the BPSness of the space-filling sources \eqref{susy3bis} are crucial in this derivation.} $\nabla^2(2A-\phi)=0$. Harmonic functions being constant on compact spaces, we set  \be e^{2A-\phi}\equiv g_s.\label{gs}\ee

As shown in Appendix \ref{sec:AppC},  imposing that the gauge-BPSness condition
still holds already constrains the allowed non-supersymmetric deformations of the pure spinor equations to\begin{align}
     &\dd_H(e^{3A-\phi}\Omega)=- i \, e^{3A- \phi}K\label{FK}\\
& \dd_H(e^{2A-\phi} {\rm Re} \, e^{i J}) =  ie^{2A- \phi} \Upsilon\\
  & \dd_H(e^{4A-\phi} {\rm Im} \, e^{i J}) =  e^{4 A} \Tilde{\ast} _6 F
\end{align}

with 
\begin{align}
K=&\frac{i}{2}\left[-t_2e^{i J}-t_1e^{-i J}+(u^1_m+p^2_m)\hat{\gamma}^m\Omega-(u^2_m+p_m^1)\Omega\hat{\gamma}^m\nonumber\right. \\
    &\left. +q^1_{mn}\hat{\gamma}^ne^{-i J}\hat{\gamma}^m+q^1_{mn}\hat{\gamma}^ne^{i J}\hat{\gamma}^m\right]
\end{align}
and 
\begin{align}
 \Upsilon=&\frac{i}{2}\left[(r_1^\ast+t_2^\ast)\Omega
    -(u^1_m+(p^2_m)^\ast)\hat{\gamma}^me^{iJ}-
    ((u^2_m)^\ast+p^1_m)e^{iJ}\hat{\gamma}^m\right.\nonumber\\
    &\left.+(q^1_{nm})^\ast\hat{\gamma}^m\Omega\hat{\gamma}^n\right]-c.c.
\end{align}
The gamma matrices $\{\hat{\gamma}^m\}$ are defined in the local vielbein, and their action on polyforms is given in Appendix \ref{sec:AppA}. Imposing our condition \eqref{condDterm} further constrains the supersymmetry breaking terms $\Upsilon$:
\begin{align}
 \Upsilon=&\frac{i}{2}\left[(r_1^\ast+t_2^\ast)\Omega+(q^1_{nm})^\ast\hat{\gamma}^m\Omega\hat{\gamma}^n\right]-c.c.
\end{align}
Expending $\Upsilon$ on the local vielbein yields
\begin{align}
    \hspace{-0.4cm}\Upsilon=&\frac{1}{2}\Big[e^A ( x_{32}-x_{23}) \Big[\Tilde{e}^1+\frac{1}{2}\Tilde{e}^1\wedge J\wedge J\Big] + e^{-A}  (x_{13} - x_{31})\Big[\Tilde{e}^2+\frac{1}{2}\Tilde{e}^2\wedge J\wedge J\Big]\\
    &+ e^{-A} (x_{21}-x_{12} )\Big[\Tilde{e}^3+\frac{1}{2}\Tilde{e}^3\wedge J\wedge J\Big]\nonumber
    + e^A  (y_{23} - y_{32})\Big[\Tilde{e}^4+\frac{1}{2}\Tilde{e}^4\wedge J\wedge J\Big]\\
    &+ e^{-A}(y_{31}-y_{13}) \Big[\Tilde{e}^5+\frac{1}{2}\Tilde{e}^5\wedge J\wedge J\Big]+ 
 e^{-A}  (y_{12} - y_{21})\Big[\Tilde{e}^6+\frac{1}{2}\Tilde{e}^6\wedge J\wedge J\Big]\Big]+X\label{ups15}
\end{align}
with $x_{ij},\ y_{ij}$ some real functions on the compact manifold, whose expressions are given in terms of the supersymmetry breaking parameters in Appendix \ref{sec:AppC}, and X a three-form specified in \eqref{Upssu3}. 

Moreover, the one-form components of $\Upsilon$ is set to zero by \eqref{gs}. The supersymmetry breaking parameters must therefore respect
\begin{align}
    x_{ij}=x_{ij}\qquad  y_{ij}=y_{ij}\qquad i,j=1,2,3.
\end{align}
It is then interesting to note that this requirements makes the five-form components of $\Upsilon$ vanish, which imposes
\be \dd(J^2)=0.\ee
We already see at this stage that the condition \eqref{condDterm} highly constrains the possible D-terms for vacua with space-filling BPS-sources.

The only remaining possibility in order to have non-vanishing D-terms is through the NSNS flux. However, the most general NSNS field-strength compatible with \eqref{condDterm} and the orientifold projection\footnote{The NSNS field-strength must be odd under the orientifold projection.} is also highly constrained
\begin{align}
    H=&\ y_{12}(e^A\Tilde{e}^1\wedge\Tilde{e}^3\wedge\Tilde{e}^4-e^{-3A}\Tilde{e}^2\wedge\Tilde{e}^3\wedge\Tilde{e}^5)\nonumber\\
    &+x_{13}(e^A\Tilde{e}^1\wedge\Tilde{e}^4\wedge\Tilde{e}^5+e^{-3A}\Tilde{e}^3\wedge\Tilde{e}^5\wedge\Tilde{e}^6)\nonumber\\
      &-x_{12}(e^A\Tilde{e}^1\wedge\Tilde{e}^4\wedge\Tilde{e}^6-e^{-3A}\Tilde{e}^2\wedge\Tilde{e}^5\wedge\Tilde{e}^6)\nonumber\\
    &-y_{13}(e^A\Tilde{e}^1\wedge\Tilde{e}^2\wedge\Tilde{e}^4+e^{-3A}\Tilde{e}^2\wedge\Tilde{e}^3\wedge\Tilde{e}^6).
\end{align}
For these backgrounds, the Bianchi identity for the NSNS flux and the B-field equation of motion \eqref{Bfieldeom} read
\be \dd H=0\qquad \dd (\Tilde{\ast}_6 H)=0,\ee
and together with the NSNS flux quantisation condition they would further constrain the possible NSNS flux, upon specifying some internal geometry. 

Finally, it is also important to note that the orientifold projection sets
\be H\wedge\Omega=0.\ee  Since the F-terms from \eqref{FK} cannot be vanishing\footnote{There
cannot be pure D-term breaking of supergravity in Minkowski space.}, this means that the breaking of supersymmetry cannot originate purely from NSNS flux components. 

Constructing ten-dimensional supergravity solutions with D-terms is difficult, and in this illustrative example we see that the mere requirement of consistency with the four-dimensional $\mathcal{N}=1$ description highly constrains the possible D-terms expression, potentially ruling out the possibility for non-vanishing D-terms for the whole class considered.

\section{Conclusion}

    In this paper, we studied non-supersymmetric solutions of type II supergravity  within the framework of Generalised Complex Geometry. The interpretation of the supersymmetry conditions in terms of calibration conditions for different types of probe D-branes led us to consider a subclass of non-supersymmetric solutions partially preserving supersymmetry, in the sense that the calibration condition for space-filling D-branes remains satisfied and such backgrounds admit only space-filling BPS sources. This calibration condition has been dubbed the gauge BPSness condition in \cite{Lust:2008zd}. On the other hand, the calibration conditions for string-like and domain-wall probe D-branes are allowed to be violated, which is encoded through the introduction of some supersymmetry breaking terms in these conditions.

    The gauge BPSness condition has been interpreted in \cite{Koerber:2007xk} as an F-term condition, making the connection with the four-dimensional $\mathcal{N}=1$ description.

    We derived a generalisation of the gauge BPSness for our class of non-supersymmetric vacua. We then investigated under which conditions the gauge BPSness can still be interpreted as an F-term condition. Interestingly, this is the case when \eqref{condDterm} is respected, i.e. when some terms violating the string-like calibration condition are set to zero. These terms belong to vector representations of the SU$(3)\times$SU$(3)$ structure, just like the modes identified with four-dimensional massive spin $\frac{3}{2}$ multiplets degrees of freedom in \cite{Grana:2009im}. Given that the violation of the string-like calibration condition has been interpreted as D-terms of the associated effective theory in \cite{Koerber:2007xk}, our condition \eqref{condDterm} restricts the possible D-terms for our class of backgrounds.

    On another note, the vanishing of these vector-like modes results in some negative semi-definite contributions to the effective potential being set to zero on-shell (the last line of the effective potential \eqref{effpotn0}). This is in agreement with the four-dimensional $\mathcal{N}=1$ picture, given that these contributions do not originate from the superpotential.

   We derived the equations of motion for this class of backgrounds, and they are significantly simpler than the ones one would derive without imposing the constraint \eqref{condDterm}, and an obvious extension of this work would be to search for such non-supersymmetric solutions.

    A subclass of these backgrounds is the one containing vacua which would dimensionally reduce to four-dimensional $\mathcal{N}=1$ supergravity solutions with non-vanishing F-terms, and vanishing D-terms, like the abundantly discussed no-scale vacua. We also presented the remarkably simple general equations of motion for such backgrounds, and one could again look for new solutions of this type.

    Finally, to illustrate this discussion we analysed how constraining it is to require \eqref{condDterm} for the class of SU$(3)$ backgrounds with space-filling BPS O5-planes. We showed that non-vanishing D-terms could only arise through NSNS flux components, while the NSNS flux expression is itself highly constrained by \eqref{condDterm}. It would then be interesting to investigate further the consequences of imposing our condition \eqref{condDterm} on different source configurations, and possibly rule out completely the possibility for D-terms in these cases, or find some new supergravity solutions with non-vanishing D-terms.

   \begin{center}
     \textbf{Acknowledgements}
\end{center}
It is a pleasure to thank Davide Cassani, Mariana Gra\~na, Luca Martucci, Michela Petrini, Alessandro Tomasiello, Thomas Van Riet and Dan Waldram for insightful discussions, and Anthony Ashmore for comments on the draft.

  \appendix
\section{Supergravity Conventions}\label{sec:AppA}
\subsection{Bosonic sector of type II supergravity in ten dimensions}

Our supergravity conventions are identical to those of \cite{Lust:2008zd}. We introduce the ten- and six-dimensional Hodge operators
\begin{align}
\Tilde{\ast}_{10}=&\ast_{10}\circ\sigma\\
\Tilde{\ast}_6=&\ast_{6}\circ\sigma
\end{align} 
with, for a $p$-form $\omega$
\begin{align}
&\ast_{10}\omega_p=-\frac{1}{p!(10-p)!}\sqrt{-g}\ \epsilon_{M_1...M_{10}}\omega^{M_{11-p}...M_{10}}\dd x^{M_1}\wedge ...\wedge \dd x^{M_{10-p}}\\
&\ast_{6}\omega_p=\frac{1}{p!(6-p)!}\sqrt{-g}\ \epsilon_{m_1...m_{6}}\omega^{m_{7-p}...m_{6}}\dd y^{m_1}\wedge ...\wedge \dd 
y^{m_{6-p}}.\end{align}
The ten-dimensional RR field-strength self duality is then
\be F^{10}=\Tilde{\ast}_{10}F^{10}.\ee
The type II pseudo-action in democratic formalism is
\be S=\frac{1}{2\kappa^2_{10}}\int \dd^{10}x\sqrt{-g}\Big\{e^{-2\phi}[R+4(\dd\phi)^2-\frac{1}{2}H^2]-\frac{1}{4}F^2\Big\}+S^{(\nn{loc})},\ee
where $2\kappa^2_{10}=(2\pi)^7\alpha'^4$ and for any real p-form $\omega$ we define $\omega^2=\omega\cdot\omega$ with $\cdot$ defined as
\be \omega\cdot\chi=\frac{1}{p!}\omega_{M_1...M_p}\chi^{M_1...M_p}.\ee
In the text, if $\omega$ is complex, then we consider $|\omega|^2=\omega\cdot\bar{\omega}$.
Varying this action and imposing the self-duality conditions, we find the following equations of motion. The dilaton equation 
    \be \nabla^2\phi-(\dd\phi)^2+\frac{1}{4}R-\frac{1}{8}H^2-\frac{1}{4}\frac{\kappa^2_{10}e^{2\phi}}{\sqrt{-g}}\frac{\delta S^{(\nn{loc})}}{\delta\phi}=0,\ee
the $B$-field equation
    \be -\dd(e^{-2\phi}\ast_{10}H)+\frac{1}{2}[\ast_{10}F\wedge F]_8+2\kappa^2_{10}\frac{\delta S^{(\nn{loc})}}{\delta B}=0,\ee
the Einstein equation 
\begin{align}
    e^{-2\phi}[&g_{MN}+2g_{MN}\dd\phi\cdot\dd\phi-2g_{MN}\nabla^2\phi+2\nabla_M\nabla_N\phi\nonumber\\
    &-
    \frac{1}{2}\iota_MH\cdot\iota_NH+\frac{1}{4}g_{MN}H\cdot H)]-\frac{1}{4}\iota_MF\cdot\iota_NF-\kappa^2_{10}T^{(\nn{loc})}_{MN}=0,
\end{align}
with \be T^{(\nn{loc})}_{MN}=-\frac{2}{\sqrt{-g}}\frac{\delta S^{(\nn{loc})}}{g^{MN}},\ee   and the RR-fluxes variation gives the Bianchi identities
\be \dd_HF=-j_\nn{source}.\ee
Combining the dilaton equation with the Einstein equations, one can write the modified Einstein equations
\begin{align}
    &R_{MN}+2\nabla_M\nabla_N\phi-\iota_MH\cdot\iota_NH-\frac{1}{4}e^{2\phi}\iota_MF\cdot\iota_NF\nonumber\\
    &-\kappa^2_{10}e^{2\phi}\Big(T^{(\nn{loc})}_{MN}+\frac{g_{MN}}{2\sqrt{-g}}\frac{\delta S^{(\nn{loc})}}{\delta\phi}\Big)=0\label{modeinstein}.
\end{align}
Finally, we define the Mukai pairing for a pair of polyforms $\omega$ and $\chi$
\be \braket{\omega,\chi}=\omega\wedge\sigma(\chi)|_6,\ee
and more generally, we use throughout the paper
\be \braket{\omega,\chi}_k=\omega\wedge\sigma(\chi)|_k,\ee where we
project onto the k-form
component of $\omega\wedge\sigma(\chi)$.

In the case of a six-dimensional manifold $\mathcal{M}$, the Mukai pairing satisfies the following property
\be \int_\mathcal{M}\braket{\dd_H\omega,\chi}=\int_\mathcal{M}\braket{\omega,\dd_H\chi}\label{Mukaid}.\ee
\subsection{Gamma matrices}
We use a real representation of the ten-dimensional gamma matrices $\Gamma_M$. The ten-dimensional chirality operator is
\be \Gamma_{(10)}=\Gamma^{01...9},\ee
with flat ten-dimensional indices. For any p-form $\omega$, we denote its image under the Clifford map $\slashed{\omega}$ by
\be \omega\equiv\frac{1}{p!}\omega_{M_1...M_p}\dd x^{M_1...M_p}\quad \longleftrightarrow\quad\slashed{\omega}=\frac{1}{p!}\omega_{M_1...M_p}\Gamma^{M_1...M_p}.\label{cliffmap}\ee

    We define the splitting of the ten-dimensional gamma matrices into four- and six-dimensional gamma matrices $\hat{\gamma}^\mu$ and $\gamma^m$ as
    \be \Gamma^\mu=e^{-A}\hat{\gamma}^\mu\otimes\mathbb{1}\qquad\Gamma^m=\gamma_{(4)}\otimes\gamma^m.\ee
The $\hat{\gamma}^\mu$ are associated to the unwarped four-dimensional metric, and $\gamma_{(4)}=i\hat{\gamma}^{0123}$ is the usual four-dimensional chirality operator. The six-dimensional chirality operator is $\gamma_{(6)}=-i\gamma^{123456}$ so we have $\Gamma_{(10)}=\gamma_{(4)}\otimes\gamma_{(10)}$. 

    The chirality of the internal spinors is then
    \be \gamma_{(6)}\eta_1=\eta_1\qquad\gamma_{(6)}\eta_2=\mp\eta_2\qquad\nn{in type IIA/IIB}.\ee
The internal gamma matrices act on a form $\omega$ as 
 \be 
 \label{gammas}
 \gamma^m\omega=(g^{mn}\iota_n+\dd y^m\wedge)\omega\qquad\omega\gamma^m=(-1)^{|\omega|+1}(g^{mn}\iota_n-\dd y^m\wedge)\omega.
 \ee
One can also define gamma matrices associated to a local internal vielbein $\{e^a\}$ 
\be 
\hat{\gamma}^a\omega=(\delta^{ab}\iota_{b}+e^a\wedge)\omega\qquad\omega\hat{\gamma}^a=(-1)^{|\omega|+1}(\delta^{ab}\iota_{b}-e^a\wedge)\omega.\label{gammavielbein}
\ee
\section{Some Generalised Complex Geometry elements}
\label{sec:AppB}
Generalised vectors are sections of the generalised tangent bundle, locally defined as $E\simeq T\mathcal{M}\oplus T^\ast\mathcal{M}$. The generalised tangent bundle admits a natural SO$(6,6)$ structure, defined by the inner product on generalised vectors
   \be (V,W)=\iota_v\rho+\iota_w\xi,\ee
    with $V=v+\xi\in\Gamma(E)$ and $W=  w+\rho\in\Gamma(E)$. Using a two-component notation for the generalised
vectors
\be V=v+\xi\equiv \begin{pmatrix}
v\\
\xi
\end{pmatrix},\ee
one can parametrise the SO$(6,6)$ generators as
\be \mathcal{O}_A=\begin{pmatrix}
\ A & 0\\
\ 0 & (A^T)^{-1}
\end{pmatrix}\qquad\mathcal{O}_b=\begin{pmatrix}
\ \mathbb{1} &\ \ 0\ \\
\ b &\ \ \mathbb{1}\ 
\end{pmatrix}\qquad\mathcal{O}_\beta=\begin{pmatrix}
\ \mathbb{1} &\ \ \beta\ \\
\ 0 &\ \ \mathbb{1}\ 
\end{pmatrix},\ee
with $A$ any GL$(d,\mathbb{R})$ matrix, $b$ any two-form and $\beta$ any two-vector.
The SO$(6,6)$ adjoint action on generalised vectors is then
\begin{align}
    V=v+\xi\quad\rightarrow\quad\mathcal{O}_A\cdot V=&\ Av+(A^T)^{-1}\xi\\
    \quad\mathcal{O}_b\cdot V=&\ v+(\xi-\iota_v b)\\
     \quad\mathcal{O}_\beta\cdot V=&\ (v-\iota_\beta \xi)+\xi.
\end{align}
    
    On another note, the generalised vectors obey the Cliff$(6,6)$ Clifford algebra
\be \{V,W\}=(V, W)\qquad V,\ W\in\Gamma(E),\ee
and the Spin$(6,6)$ spinors are sections of a spinor representation of Cliff$(6,6)$, locally isomorphic to the space of polyforms. The action of a generalised vector $V=v+\xi$ on such a spinor $\Psi$ is
\be V\cdot \Psi=\iota_v\Psi+\xi\wedge\Psi.\label{vecspinoract}\ee
A line-bundle of pure spinors is in one-to-one correspondence with an (almost) generalised complex structure, see for instance \cite{Gualtieri:2007ng} for more details.

    An almost generalised complex structure is a map
    \be \mathcal{J}\colon E\rightarrow E,\ee
respecting
\begin{align}
    \mathcal{J}^2=-\mathbb{1}\qquad\mathcal{J}^T\mathcal{I}\mathcal{J}=\mathcal{I}
\end{align}
with
\be \mathcal{I}\colon E\times E\rightarrow \mathbb{R}\quad\nn{such that}\quad\mathcal{I}(V,W)=(V, W)\qquad V,\ W\in\Gamma(E).\ee
It is then an integrable almost generalised complex structure or simply generalised complex structure if its $+i$-eigenbundle is stable under the following Courant bracket
\be [V,W]_\nn{C}=[v,w]+\mathcal{L}_v\rho-\mathcal{L}_w\xi-\frac{1}{2}\dd(\iota_v\rho-\iota_w\xi),\ee
with $V=v+\xi\in\Gamma(E)$ and $W=  w+\rho\in\Gamma(E)$. An almost generalised complex structure is $H$-integrable if its $+i$-eigenbundle is stable under the twisted Courant bracket
\be [V,W]^H_\nn{C}=[V,W]_\nn{C}+\iota_v\iota_wH.\label{twistcour}\ee
One can also define a natural action of (almost) generalised complex structures on the space of differential forms, explicitly\footnote{See \cite{Cavalcanti:2005hq} for the formal details.}:
\be \mathcal{J}\cdot=\frac{1}{2}\big(J_{mn}\dd y^m\wedge \dd y^n\wedge+2 {I^m}_n[\dd y^n\wedge,\iota_m]+P^{mn}\iota_m\iota_n\big)\label{genstructact},\ee
with
\be \mathcal{J}=\begin{pmatrix}
    I & P\\
    J & -I^T
\end{pmatrix}.\ee
\section{Supersymmetry breaking and pure spinors}\label{sec:AppC}

%\subsection{Supersymmetry breaking parameters}
We consider ten-dimensional solutions that are warped products of four-dimensional Minkowski space and an internal six-dimensional manifold with  SU$(3)\times $SU$(3)$ structure in GCG.

In  \cite{Lust:2008zd} it was shown, that the most general non-supersymmetric deformation of the pure spinor equations \eqref{ret}-\eqref{susy3} can be written as
\begin{align}
    &e^{-2A+\phi}\dd_H(e^{2A-\phi}\Psi_1)+2\dd A\wedge\nn{Re}\Psi_1-e^\phi\Tilde{\ast} _6F=\Upsilon\\
    &e^{-3A+\phi}\dd_H(e^{3A-\phi}\Psi_2)=K
\end{align}
where the the polyforms $\Upsilon$ and $K$ are decomposed on the 
 the generalised Hodge diamond with coefficients given by the supersymmetry-breaking parameters
\begin{align}
    \Upsilon=&\frac{1}{2}(-1)^{|\Psi_1|}(r_1^\ast+t_2^\ast)\Psi_2+\frac{1}{2}(-1)^{|\Psi_1|}(r_2+t_1)\bar{\Psi}_2+\frac{1}{2}(s^1_m)^\ast\gamma^m\bar{\Psi}_1+\frac{1}{2}(-1)^{|\Psi_1|}s^2_m\bar{\Psi}_1\gamma^m\nonumber\\
    &+\frac{1}{2}[u^1_m+(p^2_m)^\ast]\gamma^m\Psi_1+\frac{1}{2}(-1)^{|\Psi_1|}[(u^2_m)^\ast+p^1_m]\Psi_1\gamma^m\nonumber\\
    &+\frac{1}{2}(q^2_{mn})^\ast\gamma^m\Psi_2\gamma^n-\frac{1}{2}q^1_{mn}\gamma^n\bar{\Psi}_2\gamma^m\\
    K=&\frac{1}{2}(-1)^{|\Psi_1|}t_2\Psi_1-\frac{1}{2}(-1)^{|\Psi_1|}t_1\bar{\Psi}_1+\frac{1}{2}(u^1_m+p^2_m)\gamma^m\Psi_2+\frac{1}{2}(-1)^{|\Psi_2|}(u^2_m+p_m^1)\Psi_2\gamma^m\nonumber\\
    &+\frac{1}{2}q^1_{mn}\gamma^n\bar{\Psi}_1\gamma^m-\frac{1}{2}q^2_{mn}\gamma^m\Psi_1\gamma^n.
\end{align}

%\subsection{The D-term condition}
Imposing the gauge BPSness \eqref{susy3bis} gives
\begin{align}
    r_1+t_2=&-(r_2+t_1)\\
    s^1_m=&-[u^1_m+(p^2_m)^\ast]\\
    s^2_m=&-[u^2_m+(p^1_m)^\ast]\\
    q^1_{mn}=&\ q^2_{nm} \, .
\end{align}
Then, imposing the condition \eqref{condDterm} is equivalent to
\begin{align} u^1_m=&-\frac{1}{2}{(1+iJ_1)^k}_{m}(p^
2_k)^\ast\qquad s^1_m=-\frac{1}{2}{(1-iJ_1)^k}_{m}(p^
2_k)^\ast\label{u11}\\
u^2_m=&-\frac{1}{2}{(1+iJ_2)^k}_{m}(p^
1_k)^\ast\qquad s^2_m= -\frac{1}{2}{(1-iJ_2)^k}_{m}(p^
1_k)^\ast\, ,\label{u22}\end{align}
  where $J_{1,2}$ are the (almost) complex structures defined by $\eta_1$ and $\eta_2$
\be {{J_{1,2}}^m}_n=\frac{i}{\lVert \eta_{1,2}\rVert}\eta_{1,2}^\dagger{\gamma^m}_n\eta_{1,2}.\ee
This results in
\begin{align}
  \Upsilon=&\frac{1}{2}(-1)^{|\Psi_1|}(r_1^\ast+t_2^\ast)\Psi_2-\frac{1}{2}(-1)^{|\Psi_1|}(r_1+t_2)\bar{\Psi}_2\nonumber\\
  &+\frac{1}{2}(q^1_{nm})^\ast\gamma^m\Psi_2\gamma^n
  -\frac{1}{2}q^1_{nm}\gamma^m\bar{\Psi}_2\gamma^n\label{Ups}\\
K=&\frac{1}{2}(-1)^{|\Psi_1|}t_2\Psi_2-\frac{1}{2}(-1)^{|\Psi_1|}t_1\bar{\Psi}_2+i\nn{Im}(p^2_m)\gamma^m\Psi_2+(-1)^{|\Psi_2|}i\nn{Im}(p^1_m)\Psi_2\gamma^m\nonumber\\
    &+\frac{1}{2}q^1_{mn}\gamma^n\bar{\Psi}_1\gamma^m-\frac{1}{2}q^1_{mn}\gamma^n\Psi_1\gamma^m.\label{K}
\end{align}

The polyforms $\Theta$ and $\Xi$ in \eqref{Bth} and \eqref{Bch} then take the form
 \begin{align}
 \Theta=&\frac{i}{2}e^{A-\phi}\Big[(-1)^{|\Psi_1|}(t_2+4t_1^\ast)\Psi_1+(-1)^{|\Psi_1|}(t_1+4t_2^\ast)\bar{\Psi}_1\nonumber\\
 &-2i\nn{Im}(p^2_m)\gamma^m\Psi_2+(-1)^{|\Psi_2|}2i\nn{Im}(p^1_m)\Psi_2\gamma^m\nonumber\\
&+q^1_{mn}\gamma^n\Psi_1\gamma^m+q^1_{mn}\gamma^n\bar{\Psi}_1\gamma^m\Big]\\
  \hspace{-0.6cm}\Xi=&\frac{1}{2}e^{A-\phi}\Big[(-1)^{|\Psi_1|}(r_1^\ast+t_2^\ast)\Psi_2+(-1)^{|\Psi_1|}(r_1+t_2)\bar{\Psi}_2\nonumber\\
  &-q^{1\ast}_{mn}\gamma^n\Psi_2\gamma^m-q^1_{mn}\gamma^n\bar{\Psi}_2\gamma^m\Big].
 \end{align}
Plugging the supersymmetry breaking expansion \eqref{K} in \eqref{uR} and using \eqref{u11} and \eqref{u22} gives
\be u^1_R=u^2_R=0.\ee
\medskip
Specifying the supersymmetry breaking terms violating the D-string BPSness condition \eqref{Ups} to the case of Section \ref{sec:example} gives
\begin{align}
    \Upsilon=&\frac{1}{2}\Big[e^A (-x_{23} + x_{32}) \Tilde{e}^1 + e^{-A}  (x_{13} - x_{31})\Tilde{e}^2 + e^{-A} (-x_{12} + x_{21})\Tilde{e}^3 \nonumber\\
    &+ e^A  (y_{23} - y_{32})\Tilde{e}^4  + e^{-A}(-y_{13}+y_{31}) \Tilde{e}^5+ 
 e^{-A}  (y_{12} - y_{21})\Tilde{e}^6 \nonumber\\
 &+ 
 e^{-A} (a - x_{11} - x_{22} - x_{33}) \Tilde{e}^1\wedge\Tilde{e}^2\wedge\Tilde{e}^3 - 
 e^{-A} (y_{23} + y_{32}) \Tilde{e}^1\wedge\Tilde{e}^2\wedge\Tilde{e}^5 \nonumber\\
 &+ 
 e^{-A} (-b + y_{11} + y_{22} - y_{33}) \Tilde{e}^1\wedge\Tilde{e}^2\wedge\Tilde{e}^6 + 
 e^{-A} (b - y_{11} + y_{22} - y_{33}) \Tilde{e}^1\wedge\Tilde{e}^3\wedge\Tilde{e}^5\nonumber\\
 &+ 
 e^{-A} (y_{23} + y_{32}) \Tilde{e}^1\wedge\Tilde{e}^3\wedge\Tilde{e}^6 - 
 e^{-A} (a - x_{11} + x_{22} + x_{33}) \Tilde{e}^1\wedge\Tilde{e}^5\wedge\Tilde{e}^6 \nonumber\\
 &+ 
 e^{-A} (-b - y_{11} + y_{22} + y_{33}) \Tilde{e}^2\wedge\Tilde{e}^3\wedge\Tilde{e}^4 + (y_{12} + 
    y_{21}) (e^A \Tilde{e}^1\wedge\Tilde{e}^3\wedge\Tilde{e}^4 - 
    e^{-3A} \Tilde{e}^2\wedge\Tilde{e}^3\wedge\Tilde{e}^5)\nonumber\\
    &- (y_{13} + 
    y_{31}) (e^A \Tilde{e}^1\wedge\Tilde{e}^2\wedge\Tilde{e}^4 + 
    e^{-3A} \Tilde{e}^2\wedge\Tilde{e}^3\wedge\Tilde{e}^6) + 
 e^{-A} (a + x_{11} - x_{22} + x_{33}) \Tilde{e}^2\wedge\Tilde{e}^4\wedge\Tilde{e}^6\nonumber\\
 &- (x_{12} + 
    x_{21}) (e^A \Tilde{e}^1\wedge\Tilde{e}^4\wedge\Tilde{e}^6 - 
    e^{-3A} \Tilde{e}^2\wedge\Tilde{e}^5\wedge\Tilde{e}^6) - 
 e^{-A} (a + x_{11} + x_{22} - x_{33}) \Tilde{e}^3\wedge\Tilde{e}^4\wedge\Tilde{e}^5 \nonumber\\
 &+ (x_{23} + 
    x_{32}) (e^{-A} \Tilde{e}^2\wedge\Tilde{e}^4\wedge\Tilde{e}^5 - 
    e^{-A} \Tilde{e}^3\wedge\Tilde{e}^4\wedge\Tilde{e}^6) \nonumber\\&+ (x_{13} + 
    x_{31}) (e^A \Tilde{e}^1\wedge\Tilde{e}^4\wedge\Tilde{e}^5 + 
    e^{-3A} \Tilde{e}^3\wedge\Tilde{e}^5\wedge\Tilde{e}^6) + 
 e^{-A} (b + y_{11} + y_{22} + y_{33}) \Tilde{e}^4\wedge\Tilde{e}^5\wedge\Tilde{e}^6 \nonumber\\
 &+ 
 e^{-A} (x_{12} - x_{21}) \Tilde{e}^1\wedge\Tilde{e}^2\wedge\Tilde{e}^3\wedge\Tilde{e}^4\wedge\Tilde{e}^5 + 
 e^{-A} (x_{13} - x_{31}) \Tilde{e}^1\wedge\Tilde{e}^2\wedge\Tilde{e}^3\wedge\Tilde{e}^4\wedge\Tilde{e}^6\nonumber \\
 &+ 
 e^{-3A} (x_{23} - x_{32}) \Tilde{e}^1\wedge\Tilde{e}^2\wedge\Tilde{e}^3\wedge\Tilde{e}^5\wedge\Tilde{e}^6 + 
 e^{-A} (-y_{12} + y_{21}) \Tilde{e}^1\wedge\Tilde{e}^2\wedge\Tilde{e}^4\wedge\Tilde{e}^5\wedge\Tilde{e}^6\nonumber \\
 &+ 
 e^{-A} (-y_{13} + y_{31}) \Tilde{e}^1\wedge\Tilde{e}^3\wedge\Tilde{e}^4\wedge\Tilde{e}^5\wedge\Tilde{e}^6 \nonumber\\
 &+ 
 e^{-3A} (-y_{23} + y_{32}) \Tilde{e}^2\wedge\Tilde{e}^3\wedge\Tilde{e}^4\wedge\Tilde{e}^5\wedge\Tilde{e}^6\Big],\label{Upssu3}
\end{align}
with
\begin{align}
    x_{ij}=&\frac{1}{2}e^{2A-\phi}\text{Re}\{iq^1_{i,j+3}+iq^1_{i+3,j}-q^1_{ij}-q^1_{i+3,j+3}\}\\
    y_{ij}=&\frac{1}{2}e^{2A-\phi}\text{Im}\{+q^1_{ij}+q^1_{i+3,j+3}-iq^1_{i,j+3}-iq^1_{i+3,j}\}\\
    a=&-2e^{2A-\phi}\text{Re}(r_1+t_2)\\
    b=&2e^{2A-\phi}\text{Im}(r_1+t_2).
\end{align}
   
\bibliographystyle{JHEP}
\bibliography{biblio}

\providecommand{\href}[2]{#2}\begingroup\raggedright\begin{thebibliography}{10}

\bibitem{Komargodski:2009pc}
Z.~Komargodski and N.~Seiberg, \emph{{Comments on the Fayet-Iliopoulos Term in Field Theory and Supergravity}}, \href{https://doi.org/10.1088/1126-6708/2009/06/007}{\emph{JHEP} {\bfseries 06} (2009) 007} [\href{https://arxiv.org/abs/0904.1159}{{\ttfamily 0904.1159}}].

\bibitem{Dine:1987xk}
M.~Dine, N.~Seiberg and E.~Witten, \emph{{Fayet-Iliopoulos Terms in String Theory}}, \href{https://doi.org/10.1016/0550-3213(87)90395-6}{\emph{Nucl. Phys. B} {\bfseries 289} (1987) 589}.

\bibitem{Grana:2005sn}
M.~Grana, R.~Minasian, M.~Petrini and A.~Tomasiello, \emph{{Generalized structures of N=1 vacua}}, \href{https://doi.org/10.1088/1126-6708/2005/11/020}{\emph{JHEP} {\bfseries 11} (2005) 020} [\href{https://arxiv.org/abs/hep-th/0505212}{{\ttfamily hep-th/0505212}}].

\bibitem{Martucci:2005ht}
L.~Martucci and P.~Smyth, \emph{{Supersymmetric D-branes and calibrations on general N=1 backgrounds}}, \href{https://doi.org/10.1088/1126-6708/2005/11/048}{\emph{JHEP} {\bfseries 11} (2005) 048} [\href{https://arxiv.org/abs/hep-th/0507099}{{\ttfamily hep-th/0507099}}].

\bibitem{Lust:2008zd}
D.~Lust, F.~Marchesano, L.~Martucci and D.~Tsimpis, \emph{{Generalized non-supersymmetric flux vacua}}, \href{https://doi.org/10.1088/1126-6708/2008/11/021}{\emph{JHEP} {\bfseries 11} (2008) 021} [\href{https://arxiv.org/abs/0807.4540}{{\ttfamily 0807.4540}}].

\bibitem{Tomasiello:2007zq}
A.~Tomasiello, \emph{{Reformulating supersymmetry with a generalized Dolbeault operator}}, \href{https://doi.org/10.1088/1126-6708/2008/02/010}{\emph{JHEP} {\bfseries 02} (2008) 010} [\href{https://arxiv.org/abs/0704.2613}{{\ttfamily 0704.2613}}].

\bibitem{Koerber:2007xk}
P.~Koerber and L.~Martucci, \emph{{From ten to four and back again: How to generalize the geometry}}, \href{https://doi.org/10.1088/1126-6708/2007/08/059}{\emph{JHEP} {\bfseries 08} (2007) 059} [\href{https://arxiv.org/abs/0707.1038}{{\ttfamily 0707.1038}}].

\bibitem{Martucci:2006ij}
L.~Martucci, \emph{{D-branes on general N=1 backgrounds: Superpotentials and D-terms}}, \href{https://doi.org/10.1088/1126-6708/2006/06/033}{\emph{JHEP} {\bfseries 06} (2006) 033} [\href{https://arxiv.org/abs/hep-th/0602129}{{\ttfamily hep-th/0602129}}].

\bibitem{Grana:2009im}
M.~Grana, J.~Louis, A.~Sim and D.~Waldram, \emph{{E7(7) formulation of N=2 backgrounds}}, \href{https://doi.org/10.1088/1126-6708/2009/07/104}{\emph{JHEP} {\bfseries 07} (2009) 104} [\href{https://arxiv.org/abs/0904.2333}{{\ttfamily 0904.2333}}].

\bibitem{Grana:2011nb}
M.~Grana and F.~Orsi, \emph{{N=1 vacua in Exceptional Generalized Geometry}}, \href{https://doi.org/10.1007/JHEP08(2011)109}{\emph{JHEP} {\bfseries 08} (2011) 109} [\href{https://arxiv.org/abs/1105.4855}{{\ttfamily 1105.4855}}].

\bibitem{Bergshoeff:2001pv}
E.~Bergshoeff, R.~Kallosh, T.~Ortin, D.~Roest and A.~Van~Proeyen, \emph{{New formulations of D = 10 supersymmetry and D8 - O8 domain walls}}, \href{https://doi.org/10.1088/0264-9381/18/17/303}{\emph{Class. Quant. Grav.} {\bfseries 18} (2001) 3359} [\href{https://arxiv.org/abs/hep-th/0103233}{{\ttfamily hep-th/0103233}}].

\bibitem{Hitchin:2003cxu}
N.~Hitchin, \emph{{Generalized Calabi-Yau manifolds}}, \href{https://doi.org/10.1093/qjmath/54.3.281}{\emph{Quart. J. Math. Oxford Ser.} {\bfseries 54} (2003) 281} [\href{https://arxiv.org/abs/math/0209099}{{\ttfamily math/0209099}}].

\bibitem{Gualtieri:2007ng}
M.~Gualtieri, \emph{{Generalized complex geometry}},  \href{https://arxiv.org/abs/math/0703298}{{\ttfamily math/0703298}}.

\bibitem{Legramandi:2019ulq}
A.~Legramandi and A.~Tomasiello, \emph{{Breaking supersymmetry with pure spinors}}, \href{https://doi.org/10.1007/JHEP11(2020)098}{\emph{JHEP} {\bfseries 11} (2020) 098} [\href{https://arxiv.org/abs/1912.00001}{{\ttfamily 1912.00001}}].

\bibitem{Menet:2023rnt}
V.~Menet, \emph{{New non-supersymmetric flux vacua from generalised calibrations}}, \href{https://doi.org/10.1007/JHEP05(2024)100}{\emph{JHEP} {\bfseries 05} (2024) 100} [\href{https://arxiv.org/abs/2311.12115}{{\ttfamily 2311.12115}}].

\bibitem{Gukov:1999ya}
S.~Gukov, C.~Vafa and E.~Witten, \emph{{CFT's from Calabi-Yau four folds}}, \href{https://doi.org/10.1016/S0550-3213(00)00373-4}{\emph{Nucl. Phys. B} {\bfseries 584} (2000) 69} [\href{https://arxiv.org/abs/hep-th/9906070}{{\ttfamily hep-th/9906070}}].

\bibitem{Grana:2020hyu}
M.~Gra\~na, N.~Kovensky and A.~Retolaza, \emph{{Gaugino mass term for D-branes and Generalized Complex Geometry}}, \href{https://doi.org/10.1007/JHEP06(2020)047}{\emph{JHEP} {\bfseries 06} (2020) 047} [\href{https://arxiv.org/abs/2002.01481}{{\ttfamily 2002.01481}}].

\bibitem{Giddings:2001yu}
S.B.~Giddings, S.~Kachru and J.~Polchinski, \emph{{Hierarchies from fluxes in string compactifications}}, \href{https://doi.org/10.1103/PhysRevD.66.106006}{\emph{Phys. Rev. D} {\bfseries 66} (2002) 106006} [\href{https://arxiv.org/abs/hep-th/0105097}{{\ttfamily hep-th/0105097}}].

\bibitem{Camara:2007cz}
P.G.~Camara and M.~Grana, \emph{{No-scale supersymmetry breaking vacua and soft terms with torsion}}, \href{https://doi.org/10.1088/1126-6708/2008/02/017}{\emph{JHEP} {\bfseries 02} (2008) 017} [\href{https://arxiv.org/abs/0710.4577}{{\ttfamily 0710.4577}}].

\bibitem{Blaback:2010sj}
J.~Blaback, U.H.~Danielsson, D.~Junghans, T.~Van~Riet, T.~Wrase and M.~Zagermann, \emph{{Smeared versus localised sources in flux compactifications}}, \href{https://doi.org/10.1007/JHEP12(2010)043}{\emph{JHEP} {\bfseries 12} (2010) 043} [\href{https://arxiv.org/abs/1009.1877}{{\ttfamily 1009.1877}}].

\bibitem{Cavalcanti:2005hq}
G.R.~Cavalcanti, \emph{{New aspects of the dd**c-lemma}},  phd thesis, 1, 2005, [\href{https://arxiv.org/abs/math/0501406}{{\ttfamily math/0501406}}].

\end{thebibliography}\endgroup
\end{document}